\documentclass{aa}

\bibpunct{(}{)}{;}{a}{}{,} 
\usepackage{natbib}
\usepackage{graphicx}
\usepackage{rotating}

\usepackage{txfonts}
\def\xmm{\it XMM-Newton}
\def\igr{\it INTEGRAL}
\def\fluxcgs{~erg~cm$^{-2}$~s$^{-1}$}
\def\lumcgs{~erg~s$^{-1}$}
\def\exo{EXO~1745--248}
\def\msunyear{~M$_{\odot}$~yr$^{-1}$}

\begin{document} 
   \title{An {\xmm} and {\igr} view on the hard state of {\exo} during
     its 2015 outburst}

     \author{M. Matranga\inst{1}
          \and
           A. Papitto\inst{2}
          \and
          T. Di Salvo\inst{1}
          \and
          E. Bozzo\inst{3}
          \and
          D.~F. Torres\inst{4}
          \and
          R. Iaria\inst{1}
          \and
         L. Burderi\inst{5}
         \and
          N. Rea\inst{4,8}
          \and
           D. de~Martino\inst{6}
           \and
          C. Sanchez-Fernandez \inst{7}
          \and
          A.~F. Gambino\inst{1}
          \and
          C. Ferrigno\inst{3}
          \and 
          L. Stella\inst{2}
}
   \institute{Universit\`{a} degli Studi di Palermo, Dipartimento di Fisica e Chimica,  via Archirafi 36, 90123 Palermo, Italy
             \email{marco.matranga01@unipa.it}
\and
INAF, Osservatorio Astronomico di Roma, via di Frascati 33,I-00044, Monte Porzio Catone (Roma) , Italy
\and
ISDC Data Centre for Astrophysics, Chemin d'Ecogia 16, CH-1290 Versoix, Switzerland
\and
Institut de Ci\'{e}ncies de l’Espai (IEEC-CSIC), Campus UAB, Carrer de Can Magrans s/n, E-08193 Barcelona, Spain
\and
Universit\`{a} degli Studi di Cagliari, Dipartimento di Fisica, SP Monserrato-Sestu KM 0.7, 09042 Monserrato, Italy
\and
Osservatorio Astronomico di Capodimonte, Via Moiarello 16, 80131 Napoli, Italy 
\and
European Space Astronomy Centre (ESA/ESAC), Science Operations Department,  E-28691 Villanueva de la Ca\~{n}ada Madrid, Spain
\and
Anton Pannekoek Institute for Astronomy, University of Amsterdam, Postbus 94249, NL-1090 GE Amsterdam, the Netherlands    
}         
\abstract
  { Transient low-mass X-ray binaries (LMXBs) often show outbursts lasting typically
a few-weeks and characterized by a high X-ray luminosity ($L_{x}
\approx 10^{36}-10^{38} $~{\lumcgs}), while for most of the time they
are found in X-ray quiescence ($L_X\approx10^{31} -
10^{33}$~{\lumcgs}). {\exo} is one of them. }
   {The broad-band coverage, and the sensitivity of instrumet on board of {\xmm} and 
  {\igr}, offers the opportunity  to characterize the hard X-ray spectrum during {\exo} outburst.}
   {In this paper we report on  quasi-simultaneous  {\xmm} and {\igr} observations of the X-ray
  transient {\exo} located in the globular cluster Terzan 5, performed
  ten days after the beginning of the outburst (on 2015 March 16th) shown by the source
  between March and June 2015. The source was caught in a hard state,
  emitting a 0.8-100 keV luminosity of $\simeq10^{37}$~{\lumcgs}.}
   { The   spectral continuum was dominated by thermal Comptonization of seed
  photons with temperature $kT_{in}\simeq1.3$~keV, by a cloud with
  moderate optical depth $\tau\simeq2$ and electron temperature
  $kT_e\simeq 40$~keV. A weaker soft thermal component at temperature
  $kT_{th}\simeq0.6$--0.7~keV and compatible with a fraction of the
  neutron star radius was also detected. A rich emission line spectrum
  was observed by the EPIC-pn on-board {\xmm}; features at energies
  compatible with K-$\alpha$ transitions of ionized sulfur, argon,
  calcium and iron were detected, with a broadness compatible with
  either thermal Compton broadening or Doppler broadening in the inner parts
  of an accretion disk truncated at $20\pm6$ gravitational
  radii from the neutron star. Strikingly, at least one narrow
  emission line ascribed to neutral or mildly ionized iron is needed
  to model the prominent emission complex detected between 5.5 and
  7.5~keV. The different ionization state and broadness suggest an
  origin in a region located farther from the neutron star than where
  the other emission lines are produced.  Seven consecutive type-I
  bursts were detected during the {\xmm} observation, none of which
  showed hints of photospheric radius expansion.  A thorough search 
  for coherent pulsations from the EPIC-pn light curve did not result 
  in any significant detection. Upper
  limits ranging from a few to 15\% on the signal amplitude were set,
  depending on the unknown spin and orbital parameters of the system.}
   {}

   \keywords{Techniques: spectroscopic -- Stars: neutron -- X-rays: binaries -- X-rays: bursts -- X-rays: individuals: {\exo}}

   \maketitle
%

\section{Introduction}


Globular clusters are ideal sites for the formation of binary systems
hosting a compact object thanks to the frequent dynamical interaction
caused by their dense environment \citep{1997A&ARv...8....1M}. Low mass
X-ray binaries (LMXB) formed by a neutron star (NS) that accretes
matter lost by a companion, low mass star are particularly favored, as
stellar encounters may cause the lower mass star of a binary to be
replaced by an heavier NS \citep{Verbunt.etal:87}. Some of the densest
and most massive globular clusters have the highest predicted rates of
stellar interactions and host a numerous population of LMXBs 
\citep{Heinke.etal:03b}.

Terzan 5 is a compact, massive cluster at a distance of 5.5 kpc 
\citep{2007A&A...470.1043O} which hosts at least three
stellar populations with different iron abundances; the observed
chemical pattern suggests that it was much more massive in the past,
so to be able to hold the iron rich ejecta of past supernova
explosions \citep{2009Natur.462..483F,2013ApJ...779L...5O},
and \citep[][]{ferraro2016}. Terzan 5
has the highest stellar interaction rate than any cluster in the
Galaxy \citep{Verbunt.etal:87,Heinke.etal:03a,Bahramian:2013}. 
This reflects into the largest population known of millisecond radio pulsars
\citep[34;][]{Ransom.etal:05, Hessels.etal:06}, and in at least 50
X-ray sources, including a dozen likely quiescent LMXBs
\citep{Heinke.etal:06b}. The populations of millisecond radio pulsars
and LMXBs are linked from an evolutionary point of view, as mass
accretion in a LMXB is expected to speed up the rotation of a NS down
to a spin period of a few milliseconds \citep{alpar1982}. This link
was confirmed by the discovery of accreting millisecond pulsars
\citep[AMSPs;][]{wijnands1998}, and by the observations of binary
millisecond pulsars swinging between a radio pulsar and an accretion
disk state on time scales that can be as short as weeks
\citep{archibald2009,Papitto.etal:13,Bassa.etal:14}. Globular clusters like Terzan 5
are preferential laboratories to study the relation between these two
classes of sources.

Many LMXBs are X-ray transients; they show outbursts lasting typically
a few-weeks and characterized by a high X-ray luminosity ($L_{x}
\approx 10^{36}-10^{38} $~{\lumcgs}), while for most of the time they
are found in X-ray quiescence ($L_X\approx10^{31} -
10^{33}$~{\lumcgs}). X-ray transient activity has been frequently
observed from Terzan 5 since 1980s
\citep{Makishima.etal:81,Warwick.etal:88,Verbunt.etal:95} and ten
outbursts have been detected ever since \citep[see, e.g.,  Table~1
  in][]{degenaar2012}. The large number of possible counterparts in
the cluster complicates the identification of the transient
responsible for each event when a high spatial resolution X-ray (or
radio) observation was not available. As a consequence, only three
X-ray transients of Terzan 5 have been securely identified, {\exo}
\citep[Terzan 5 X--1, active in 2000, 2011 and
  2015][]{Makishima.etal:81,2000IAUC.7454....1M,Heinke.etal:03a,2012PASJ...64...91S,tetarenko2016},
IGR J17480--2446 \citep[Terzan 5 X--2, active in
  2010;][]{2011A&A...526L...3P,2011MNRAS.414.1508M} and Swift
J174805.3--244637 \citep[Terzan 5 X--3, active in
  2012;][]{2014ApJ...780..127B}.





The first confirmed outburst observed from {\exo} took place in 2000,
when a \emph{Chandra} observation could pin down the location of the
X-ray transient with a sub-arcsecond accuracy
\citep{Heinke.etal:03a}. The outburst lasted $\sim100$~d, showing a
peak of X-ray luminosity\footnote{Throughout this paper we evaluate
  luminosities and radii for a distance of 5.5 kpc, which was
  estimated by \citet{2007A&A...470.1043O} with an uncertainty of 0.9
  kpc. There is also an determination from \citet{Valenti.etal:07} for the distance (5.9kpc) consistent within errors with Ortolani's distance.} $\sim6\times10^{37}$~{\lumcgs} \citep[][]{degenaar2012}. The
X-ray spectrum was dominated by thermal Comptonization in a cloud with
a temperature ranging between a few and tens of keV
\citep{Heinke.etal:03a,Kuulkers.etal:03}; a thermal component at
energies of $\approx 1$ keV, and a strong emission line at energies
compatible with the Fe K-$\alpha$ transition were also present in the
spectrum. More than 20 type-I X-ray bursts were observed, in none of
which burst oscillations could be detected \citep{galloway2008}. Two
of these bursts showed evidence of photospheric radius expansion, and
were considered by \citet{ozel2009} to draw constraints on the mass
and radius of the NS. A second outburst was observed from {\exo} in
2011, following the detection of a superburst characterized by a decay
timescale of $\approx 10$ hr \citep{2012MNRAS.426..927A,
  Serino.etal:12}. The outburst lasted $\approx20$~d, reaching an
X-ray luminosity of $9\times10^{36}$~{\lumcgs}. \citet{degenaar2012}
found a strong variability of the X-ray emission observed during
quiescence between the 2000 and the 2011 outburst, possibly caused by
low-level residual accretion.

A new outburst from Terzan 5 was detected on 2015 March, 13
\citep{2015ATel.7240....1A}. It was associated to {\exo} based on the
coincidence between its position \citep{Heinke.etal:06b} and the
location of the X-ray source observed by Swift XRT
\citep{2015ATel.7247....1L} and of the radio counterpart detected by
the Karl G. Jansky Very Large Array
\citep[VLA;][]{2015ATel.7262....1T}, with an accuracy of 2.2 and 0.4
arcsec, respectively. The outburst lasted $\approx100$ d and attained
a peak X-ray luminosity of $10^{38}$~{\lumcgs}, roughly a month into
the outburst \citep{tetarenko2016}. The source performed a transition
from a hard state (characterized by an X-ray spectrum described by a
power law with photon index ranging from 0.9 to 1.3) to a soft state
(in which the spectrum was thermal with temperature of
$\approx2$--$3$~keV) a few days before reaching the peak flux
\citep{2015ATel.7430....1Y}. The source transitioned back to the hard
state close to the end of the outburst. \citet{tetarenko2016} showed
that throughout the outburst the radio and X-ray luminosity correlated
as $L_R \propto L_X^{\beta}$ with $\beta=1.68^{+0.10}_{-0.09}$,
indicating a link between the compact jet traced by the radio emission
and the accretion flow traced by the X-ray output. The optical
counterpart was identified by \citet{ferraro2015}, who detected the
optical brightening associated to the outburst onset in {\it Hubble
  Space Telescope} images; the location of the companion star in the
color-magnitude diagram of Terzan 5 is consistent with the main
sequence turn-off. We stress that the HST study suggests that {\exo} 
is in an early phase of accretion stage with the donor expanding and
filling its Roche lobe thus representing a prenatal stage of a millisecond 
pulsar binary. This would make more interesting  the study of this 
source as well as linking what we stated above regarding rotation-powered 
MSPs and AMSPs.

Here we present an analysis of the X-ray properties of {\exo}, based
on an {\xmm} observation performed $\approx10$ days into the
2015 outburst, when the source was in the hard state. 

The main goal of this paper is to observe at a better statistics the region of spectrum
around the iron line. Then adding the  broad-band coverage allowed by {\igr} observations,
we are able to study the possible associated reflection features and give a definite 
answer on the origin of the iron line.
We also make use of additional monitoring observations of the source 
carried out with {\igr} during its 2015 outburst to spectroscopically
confirm the hard-to-soft spectral state transition displayed by EXO~1745-248 
around 57131~MJD \citep[as previously reported by][]{tetarenko2016}.
We stress out that this transition was observed by {\it Swift}. In order to
understand the physical properties of this state, we performed an observation
with {\xmm } allowing more sensitive and higher resolution data.
We focus in Sec.~\ref{sec:spectrum} on the shape of the X-ray spectrum and in
Sec.~\ref{sec:timing} on the properties of the temporal variability, while
an analysis of the X-ray bursts observed during the considered
observations is presented in Sec.~\ref{sec:burst}.


\section{Observations and Data Reduction}
\label{sec:reduc} 

\subsection{XMM-Newton}

{\xmm} observed {\exo} for 80.8 ks starting on 2015, March 22 at 04:52
(UTC; ObsId 0744170201). Data were reduced using the SAS (Science
Analysis Software) v.14.0.0.

The EPIC-pn camera observed the source in timing mode to achieve a
high temporal resolution of 29.5~$\mu$s and to limit the effects of
pile-up distortion of the spectral response during observations of
relatively bright Galactic X-ray sources. A thin optical blocking
filter was used. In timing mode the imaging capabilities along one of
the axis are lost to allow a faster readout. The maximum number of
counts fell on the RAWX coordinates 36 and 37. To extract the source
photons we then considered a 21 pixel-wide strip extending from
RAWX=26 to 46. Background photons were instead extracted in the region ranging
from RAWX=2 to RAWX=6. Single and double events were retained. Seven
type-I X-ray bursts took place during the {\xmm} observation with a
typical rise time of less than 5~s and a decay e-folding time scale
ranging from 10 to 23~s. In order to analyze the {\it persistent}
(i.e. non-bursting) emission of {\exo} we identified the start time of
each burst as the first 1 s-long bin that exceeded the average
count-rate by more than 100~ s$^{-1}$, and removed from the analysis a
time interval spanning from 15~s before and 200~s after the burst
onset. After the removal of the burst emission, the mean count rate
observed by the EPIC-pn was 98.1~s$^{-1}$. Pile-up was not expected to
affect significantly the spectral response of the EPIC-pn at the
observed {\it persistent} count rate (Guainazzi et al. 2014; Smith et
al. 2016)\footnote{http://xmm2.esac.esa.int/docs/documents/CAL-TN-0083.pdf,
  http://xmm2.esac.esa.int/docs/documents/CAL-TN-0018.pdf}. To check
the absence of strong distortion we run the SAS task {\it epatplot},
and obtained that the fraction of single and double pattern events
falling in the 2.4--10 keV band were compatible with the expected value
within the uncertainties. Therefore, no pile-up correction method was
employed. The spectrum was re-binned so to have not more than three
bins per spectral resolution element, and at least 25 counts per
channel.


The MOS-1 and MOS-2 cameras were operated in Large Window and Timing
mode, respectively. At the count rate observed from {\exo} both
cameras suffered from pile up at a fraction exceeding $10\%$ and were
therefore discarded for further analysis.

We also considered data observed by the Reflection Grating
Spectrometer (RGS), which operated in Standard Spectroscopy mode. We
considered photons falling in the first order of diffraction. The same
time filters of the EPIC-pn data analysis were applied.

\section{INTEGRAL} 
\label{sec:integral}  

We analyzed all {\igr} \citep{w03} available data collected in the direction 
of  {\exo} during the source outburst in 2015. These observations included both
publicly available data and  our proprietary data in AO12 cycle.

The reduction of the {\igr} data was performed using the standard Offline 
Science Analysis (OSA) version 10.2 distributed by the ISDC \citep{courvoisier03}. 
{\igr} data are divided into science windows (SCW), i.e. different pointings 
lasting  each $\sim 2-3$\,ks. We analyzed data from the IBIS/ISGRI \citep{ubertini03,lebrun03}, 
covering the energy range 20-300~keV energy band, and from the two 
JEM-X monitors \citep{lund03}, operating in the range 3-20~keV. 
As the source position varied with respect to the aim point of the 
satellite during the observational period ranging from 2015 March 12 at 19:07 
(satellite revolution 1517) to 2015 April 28 at 11:40 UTC (satellite 
revolution 1535), the coverage provided by IBIS/ISGRI was generally much 
larger than that of the two JEM-X monitors due to their smaller field of view.  

\begin{figure}
\centering
   \includegraphics[width=6cm,angle=-90]{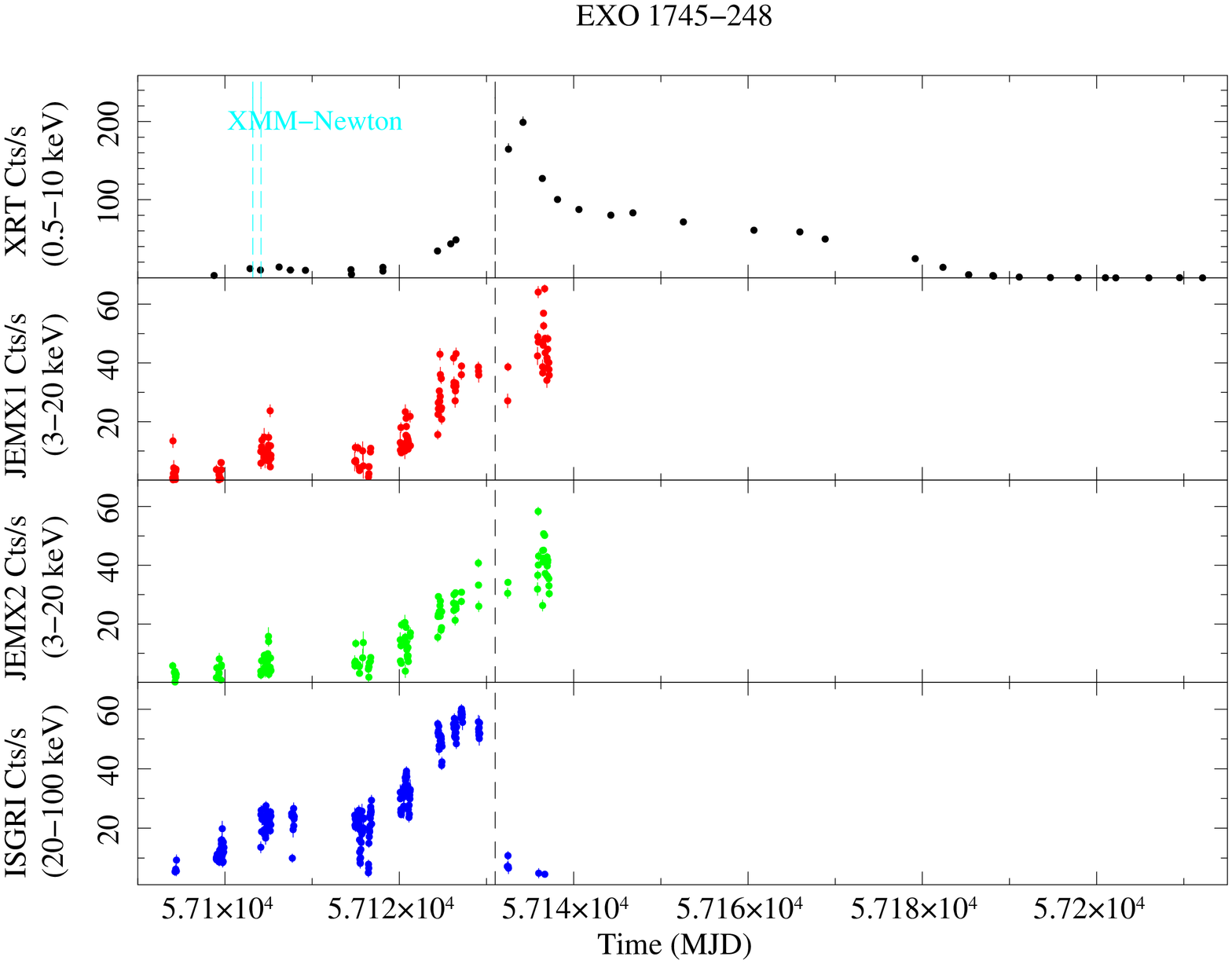}
  \caption{Lightcurve of the 2015 outburst displayed by {\exo} as observed by  IBIS/ISGRI and JEM-X on-board {\igr}. 
   For completeness, we report also the lightcurve 
  obtained from Swift/XRT and published previously by \citet{tetarenko2016}. 
  The hard-to-soft spectral state transition of {\exo} around 57131~MJD discussed by \citet{tetarenko2016} is marked with a dashed vertical line in the above plots (around this date the 
  count-rate of the source in the IBIS/ISGRI decreases significantly, while it
  continues to raise in JEM-X).}
  \label{fig:intlc}
\end{figure}

As the source was relatively bright during the outburst, we extracted a lightcurve
with the resolution of 1 SCW for both IBIS/ISGRI and the two JEM-X units. 
This is shown in Fig.~\ref{fig:intlc}, together with the monitoring observations
provided by Swift/XRT (0.5-10~keV). The latter data were retrieved from the 
Leicester University on-line analysis tool \citep{evans09} and used only to 
compare the monitoring provided by the Swift and {\igr} satellites. 
We refer the reader to \citet{tetarenko2016} for more details on the Swift 
data and the corresponding analysis. In agreement with the results discussed 
by these authors, also the {\igr} data show that the source underwent a 
hard-to-soft spectral state transition around 57131~MJD. In order to prove 
this spectral state change more quantitatively, we extracted two sets 
of {\igr} spectra accumulating all data before and after this date for 
ISGRI, JEM-X1, and JEM-X2. 
Analysis of broad-band {\igr} spectra for both hard and soft state is reported 
in Sect. \ref{sec:integral_spec}. 

We also extracted the ISGRI and JEM-X data by using only the observations 
carried out during the satellite revolution 1521, as the latter partly overlapped 
with the time of the {\xmm} observation. The broad-band fit of the combined 
quasi-simultaneous {\xmm} and {\igr} spectrum of the source is discussed in Sect.~3.2. 

We removed from the data used to extract all JEM-X and ISGRI spectra mentioned 
above the SCWs corresponding to the thermonuclear bursts detected by {\igr}. 
These were searched for by using the JEM-X lightcurves collected with 2~s 
resolution in the 3-20~keV energy band. A total of 4 bursts were clearly detected
by JEM-X in the SCW 76 of revolution 1517 and in the SCWs 78, 84, 94 of 
revolution 1521. The onset times of these bursts were 57094.24535~MJD, 57104.86423~MJD, 
57104.99993~MJD, and 57105.25787~MJD, respectively. None of these bursts 
were significantly detected by ISGRI or showed evidence for a photospheric
radius expansion. Given the limited statistics of the two JEM-X monitors during 
the bursts we did not perform any refined analysis of these events. 
 
\begin{figure}
\centering
   \includegraphics[width=6cm,angle=-90]{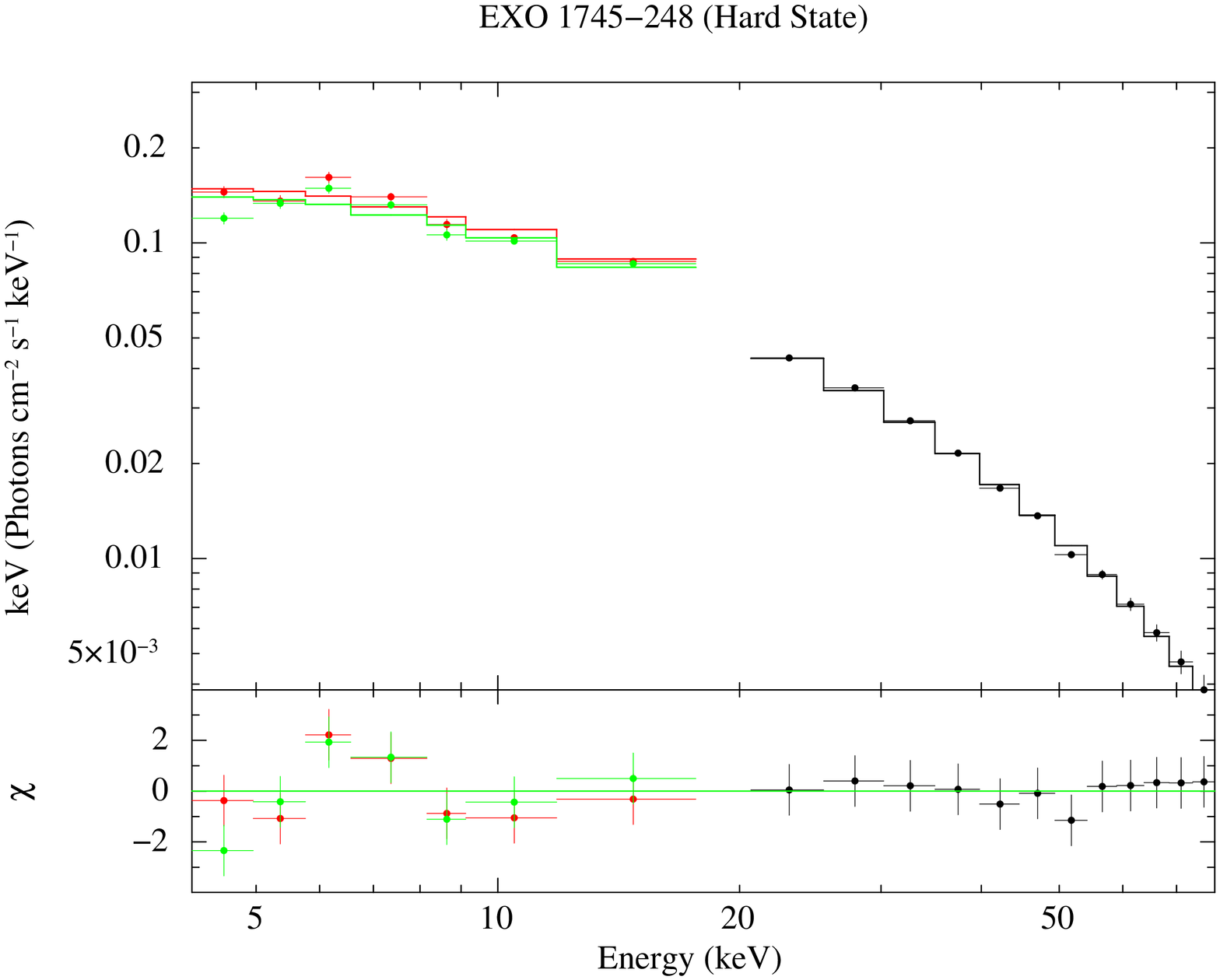}
   \includegraphics[width=6cm,angle=-90]{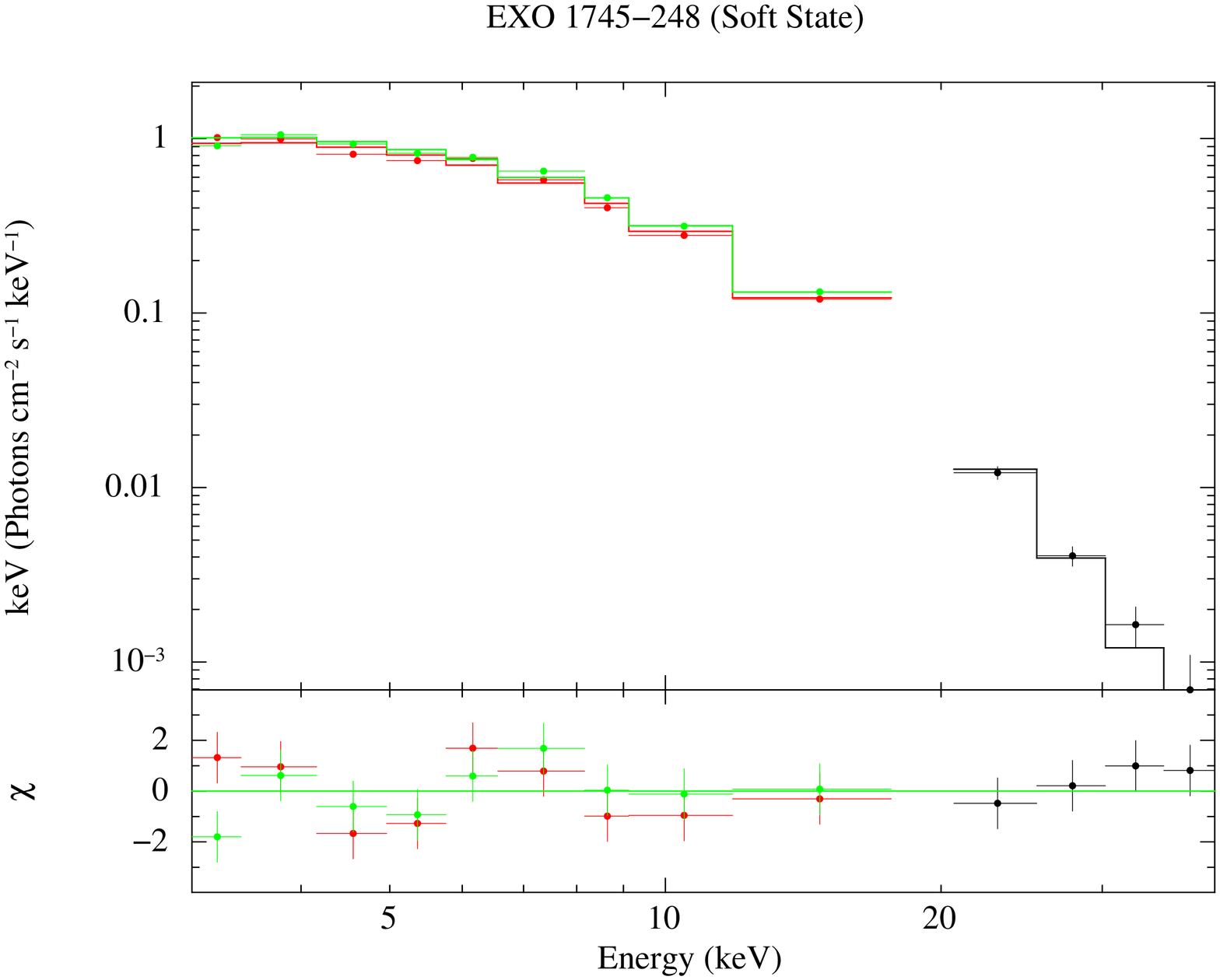}
  \caption{The broad-band spectrum of {\exo} as observed by {\igr} in the hard (top)
  and soft (bottom) state  (ISGRI data are in black, JEM-X1 data in red, and JEM-X2 
  data in green). For both states the best fit to the spectrum was obtained with 
  an absorbed cut-off power-law model (see text for details). 
  The residuals from the best fits are shown in the bottom panels of the upper and 
  lower figure.} 
  \label{fig:intspe}
\end{figure}

\section{Spectral Analysis}
\label{sec:spectrum}

Spectral analysis has been performed using XSPEC v.12.8.1
\citep{Arnaud:96}.  
Interstellar absorption was described by the TBAbs component.
For each fit we have used element abundances from \citet{2000ApJ...542..914W}. 
The uncertainties on the parameters quoted in the following are evaluated
at a 90\% confidence level. 

\subsection{Hard and soft INTEGRAL spectra}
\label{sec:integral_spec} 

The broad-band {\igr} spectrum of the 
source could be well described by using a simple absorbed power-law model 
with a cut-off at the higher energies (we fixed in all fits the absorption 
column density to the value measured by {\xmm} in Sect.~3.2, 
i.e. $N_{\rm H}$ = 2.02$\times$10$^{22}~$ cm$^{-2}$). 
In the hard state ($\chi^2_{\rm red}/d.o.f.=1.2/21$), we measured a 
power-law photon index $\Gamma$=1.1$\pm$0.1 and a cut-off energy of 
23$\pm$2~keV. 
The source X-ray flux was (2.9$\pm$0.2)$\times$10$^{-9}$~erg~cm$^{-2}$~s$^{-1}$
in the 3-20~keV energy band, (1.0$\pm$0.1)$\times$10$^{-9}$~erg~cm$^{-2}$~s$^{-1}$ 
in the 20-40~keV energy band, and (6.1$\pm$0.3)$\times$10$^{-10}$~erg~cm$^{-2}$~s$^{-1}$
in the 40-100~keV energy band. The effective exposure time was of 123~ks 
for ISGRI and 75~ks for the two JEM-X units. 
In the soft state ($\chi^2_{\rm red}/d.o.f.=1.3/17$), we measured a power-law
photon index $\Gamma$=0.6$\pm$0.2 and a cut-off energy of 3.8$\pm$0.5~keV.
The source X-ray flux was (9.5$\pm$0.5)$\times$10$^{-9}$~erg~cm$^{-2}$~s$^{-1}$
in the 3-20~keV energy band, (1.6$\pm$0.3)$\times$10$^{-10}$~erg~cm$^{-2}$~s$^{-1}$
in the 20-40~keV energy band, and (1.1$\pm$0.5)$\times$10$^{-12}$~erg~cm$^{-2}$~s$^{-1}$
in the 40-100~keV energy band. The effective exposure time was of 32~ks for 
ISGRI and 20~ks for the two JEM-X units. 
The two broad-band spectra and the residuals from the best fits are shown 
in Fig.~\ref{fig:intspe}.

\subsection{The 2.4--10 keV EPIC-pn spectrum}
\label{sec:epnspec} 

\begin{figure}
 \resizebox{\hsize}{!} 
{\includegraphics{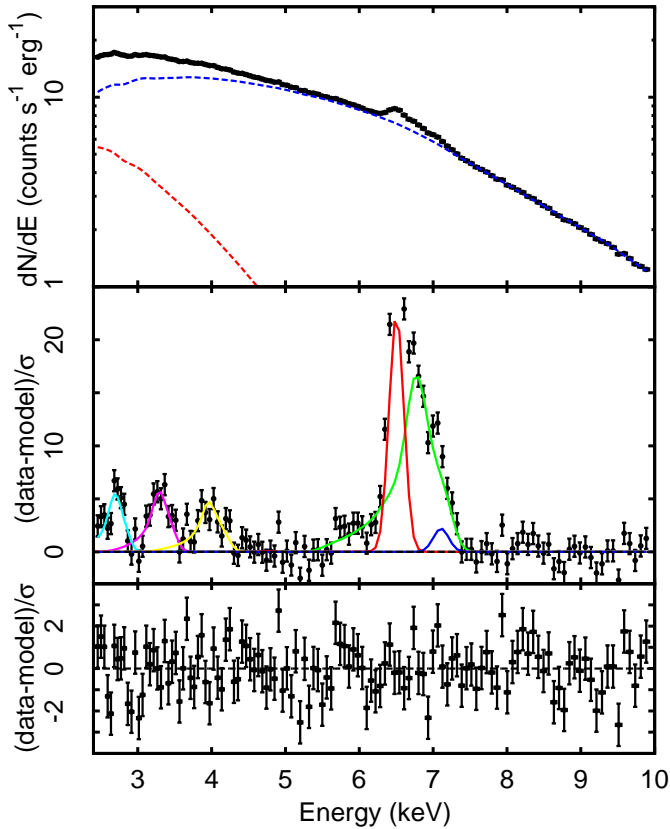}}

\caption{Spectrum observed by the EPIC-pn between 2.4 and 10 keV
  together with the best fitting black body (red dashed line) and
  Comptonization (blue dashed line) component of Model II listed in
  Table~\ref{tab:epn} (top panel). Residuals obtained when the six
  emission features at energies $E_1=6.75$ (green solid line),
  $E_2=6.48$ (red solid line), $E_3=7.12$ (blue solid line),
  $E_4=2.74$ (cyan), $E_5=3.30$ (magenta), $E_6=3.94$ (yellow) are
  removed from Model II (middle panel). The model is not fitted after
  the line removal, so the residuals are plotted for an illustrative
  purpose, only. Residuals left by Model II are plotted in the bottom
  panel.}
    \label{fig:epn}
\end{figure}

We first considered the spectrum observed by the EPIC-pn at energies
between 2.4 and 10 keV (see top panel of Fig.~\ref{fig:epn}), as a
soft-excess probably related to uncertainties in the redistribution
calibration affected data taken at lower energies (see the discussion
in Guainazzi et
al. 2015\footnote{http://xmm2.esac.esa.int/docs/documents/CAL-TN-0083.pdf},
and references therein). Interstellar absorption was described by the
TBAbs component \citep{2000ApJ...542..914W} the 
photoelectric cross sections from \citet{Verner.etal:96} with the hydrogen column
density fixed to $N_H=2\times10^{22}$~cm$^{-2}$, as indicated by the
analysis performed with the inclusion of RGS, low energy data (see
Sec.~\ref{sec:broad}). The spectral continuum was dominated by a hard,
power-law like component with spectral index $\Gamma\simeq2$, which we
modeled as thermal Comptonization of soft photons with
$kT_{in}\simeq1.3$~keV, by using the model \textsc{nthcomp}
\citep{1996MNRAS.283..193Z,1999MNRAS.309..561Z}. As the electron
temperature fell beyond the energy range covered by the EPIC-pn, we
fixed such parameter to 37~keV, as suggested by the analysis of data
taken by INTEGRAL at higher energies (see Sec.~\ref{sec:broad}). We
modeled the strong residuals left by the Comptonization model at low
energies with a black-body with effective temperature
$kT_{th}\simeq0.6$ keV and emission radius
$R_{th}\simeq5.5\,d_{5.5}$~km, where $d_{5.5}$ is the distance to the
source in units of 5.5~kpc. The addition of such a component was
highly significant as it decreased the model reduced chi-squared from
47.9 to 26.5 for the two degrees of freedom less, out of 122.
The F-test gives a probability of chance improvement of $\sim 1.3\times10^{-15}$
for the addition of the blackbody component. In this case the use of the F-test is justified by the fact that the 
model is additive \citep[see][]{Orlandini.etal:12}.

Even after the addition of a thermal component, the quality of the
spectral fit was still very poor mainly because of residuals observed
at energies of the Fe K-$\alpha$ transition (6.4--7 keV; see middle
panel of Fig.~\ref{fig:epn}). The shape of this emission complex is
highly structured and one emission line was not sufficient to provide
an acceptable modeling. We then modeled the iron complex using three
Gaussian features centered at energies
$E_1\simeq6.75^{+0.02}_{-0.03}$, $E_2\simeq6.48^{+0.03}_{-0.01}$ and
$E_3\simeq7.12^{+0.04}_{-0.07}$ keV. These energies are compatible
with K-$\alpha$ transition of ionized Fe XXV, K-$\alpha$ and K-$\beta$
of neutral or weakly ionized Fe (I-XX), respectively. The ionized iron
line is relatively broad ($\sigma_1=0.24\pm0.03$~keV) and strong
(equivalent width $EW_1=62.0\pm0.02$~eV), while the others are weaker
and have a width lower than the spectral resolution of the
instrument. In order to avoid correlation among the fitting
parameters, we fixed the normalization of the K-$\beta$ transition of
weakly ionized iron to one tenth of the K-$\alpha$. The addition of
the three Fe emission lines decreased the model $\chi^2$ to 266 for
114 degrees of freedom. In this case, for the addition of the 
iron lines, the F-test probability results to be $\sim 2.7\times10^{-16}$.
Three more emission lines were required at
lower energies, $E_4=2.74^{+0.01}_{-0.03}$, $E_5=3.30(3)$ and
$E_6=3.94^{+0.05}_{-0.06}$ keV, compatible with K-$\alpha$ transitions
of S XVI, Ar XVIII, and Ca XX (or XIX), respectively. The significance
of these lines has been evaluated with an F-test, giving probabilities
of $3\times10^{-4}$, $9\times10^{-6}$ and $7.7\times10^{-8}$,
respectively, that the improvement of the fit $\chi^2$ obtained after
the addition of the line is due to chance. The chi-squared of the
model (dubbed Model I in Table~\ref{tab:epn}) is $\chi^2=154.5$ for
106 degrees of freedom.



\footnotesize 


\begin{table*}[!h]
\caption{
Best fit parameters for the models used to fit the spectrum of EXO 1745--248. 
In details, Model I is given by tbabs * (bbody + Gaussian(E${_1}$) + Gaussian(E${_2}$) + Gaussian(E${_3}$) + Gaussian(E${_4}$) + Gaussian(E${_5}$) + Gaussian(E${_6}$) + nthComp) - Model II: tbabs * (bbody + Diskline(E${_1}$) + Gaussian(E${_2}$) + Gaussian(E${_3}$) + Diskline(E${_4}$) + Diskline(E${_5}$) + Diskline(E${_6}$)+nthComp) - Model II$*$: the same as Model II but applied to the broad band spectrum
- Model III: tbabs * (bbody + Diskline(E${_1}$) + Gaussian(E${_2}$) + Gaussian(E${_3}$) + Diskline(E${_4}$) + Diskline(E${_5}$) + rdblur*rfxconv*nthComp) - Model IV: tbabs * (bbody + Diskline(E${_1}$) + Gaussian(E${_2}$) + Gaussian(E${_3}$) + Diskline(E${_4}$) + Diskline(E${_5}$) + Diskline(E${_6}$) + rdblur*pexriv+nthComp). Where tbabs describes the photoelectric absorption, bbody is the blackbody emission and nthComp describes the primary Comptonization spectrum. The component rdblur describes the smearing effects, while rfxconv, pexriv  are different reflection models (see text for details).  }           
     
\centering          

\begin{tabular}{clccccc}     
\hline\hline 
 & & \multicolumn{2}{c}{EPIC-pn (2.4--11~keV)} & \multicolumn{3}{c}{Broadband (0.35--180~keV)}\\
Component & Parameter & Model I & Model II & Model II* & Model III & Model IV  \\

\hline

tbabs &N$_{H}\,$($\times10^{22}$ cm$^{-2}$) & 
(2.0) & (2.0) & $2.02\pm0.04$ & $2.13\pm0.05$ & $2.06\pm0.05$ \\
\noalign{\medskip}
 
bbody &$kT_{th}$ (keV) & 0.58$^{+0.03}_{-0.06}$ & 0.64$^{+0.04}_{-0.02}$ &
0.63$\pm$0.04&$0.73\pm0.03$ \\

bbody &R$_{bb}\,$ ($d_{5.5}$~km) &  
5.5$^{+0.8}_{-0.4}$ & $4.6\pm0.2$ & $4.5\pm0.5$ & $3.8\pm0.2$ &  $4.4\pm0.4$ \\
\noalign{\medskip}

nthComp &$\Gamma$ & 2.06$^{+0.08}_{-0.12}$ & $2.02^{+0.19}_{-0.09}$& $1.93\pm0.07$& $1.89\pm0.08$& $1.90\pm0.05$\\
nthComp &$kT_{e}$ (keV) & (37.0) & (37.0)& 37.2$^{+6.9}_{-5.1}$& $40^{+7}_{-5}$&33.6$^{+5.7}_{-4.4}$\\
nthComp &$kT_{in}$ (keV) & 1.33$^{+0.06}_{-0.14}$ & $1.3\pm0.1$& $1.27\pm0.06$&$1.34\pm0.07$&1.25$^{+0.08}_{-0.04}$\\
nthComp &R$_{w}$ ($d_{5.5}$~km)  &  $1.6\pm0.3$ & $1.5\pm0.3$ & $2.8\pm0.3$ & $2.4\pm0.3$ &  $2.5\pm0.4$\\
nthComp & $F_{Comp}$  & $0.86\pm0.02$ &$0.86^{+0.03}_{-0.02}$ &$0.96\pm0.08$ &$0.85\pm0.08$ & $0.83\pm0.08$\\
\noalign{\medskip}

Diskl/rdblur &$\beta_{\mbox irr}$ & ... & $-2.44^{+0.04}_{-0.06}$ & $-2.44\pm0.07$& $-2.24\pm0.07$ & $-2.43\pm0.05$ \\
Diskl/rdblur &$R_{in}$ ($R_g$) & ... & $20^{+4}_{-6}$ & $20\pm6$& $<8.5$& $18.3^{+3.9}_{-6.2}$\\
Diskl/rdblur &$R_{out}$ ($R_g$) & ... & $(10^7)$ &$(10^7)$& $(10^7)$& $(10^7)$\\
Diskl/rdblur &$i$ ($^{\circ}$) & ... & $37^{+2}_{-3}$ & $37\pm3$& $38\pm1$& $37.2^{+2.1}_{-1.7}$  \\
\noalign{\medskip}
reflection &$\Omega_r/2\pi$ & ... & ... & ... & $0.22\pm0.04$ \\
reflection &$log\xi$ & ... & ... & ... & $2.70\pm0.07$ & $2.39^{+0.41}_{-0.27}$\\
reflection &$T_{disk}$ (k) & ... & ... & ... & ... & $(10^6)$ \\
reflection & Norm ($\times 10^{-2})$ & ... & ... & ... & ... & $0.86\pm0.55$ \\


\noalign{\medskip}
Gauss/Diskl &$E_1$ (keV) &$6.75^{+0.02}_{-0.03}$ & $6.75\pm0.02$ & $6.74\pm0.02$& ...& $6.75\pm0.02$ \\
Gauss/Diskl &$\sigma_1$ (keV) &$0.24^{+0.03}_{-0.02}$ & ... & ... & ... \\
Gauss/Diskl &$N_1$  &$6.0^{+0.7}_{-0.5}$ & $6.6^{+0.6}_{-0.4}$ & $7.1\pm0.1$& ...& $6.7^{+0.2}_{-0.4}$\\
Gauss/Diskl &${EW}_1$ (eV)&$62.0\pm0.02$  & $68.2\pm0.04$ & $72.9\pm2.5$ & ... & $68.6\pm2.4$ \\
\noalign{\medskip}

Gaussian &$E_2$ (keV) & $6.48^{+0.03}_{-0.01}$ & $6.50\pm0.01$ & $6.50\pm0.02$ & $6.49\pm0.02$& $6.49\pm0.02$\\
Gaussian &$\sigma_2$ (keV) & (0.0) & (0.0) & (0.0)& (0.0)& (0.0)\\
Gaussian &$N_2$  & $2.8\pm0.3$& $3.2\pm0.2$ & $3.2\pm0.2$& $2.4\pm0.2$& $3.2\pm0.3$\\
Gaussian &${EW}_2$ (eV)&$26.8\pm0.02$  & $31.6\pm0.2$ &  $31.3\pm1.4$&  $23.1\pm1.9$& $31.1\pm1.9$ \\
\noalign{\medskip}

Gaussian &$E_3$ (keV) &$7.12_{-0.07}^{+0.04}$ & $7.09\pm0.07$& (7.06) &  (7.06)&  (7.06) \\
Gaussian &$\sigma_3$ (keV) & (0.0) & (0.0) & (0.0) & (0.0) & (0.0)\\
Gaussian &$N_3$  & ($N_2/10$) & ($N_2/10$)& ($N_2/10$)& ($N_2/10$)& ($N_2/10$)\\
Gaussian &${EW}_3$ (eV)&$3.1\pm0.1$  & $3.6\pm0.1$ &$3.5\pm0.7$ &$2.7\pm0.8$& $3.5\pm0.9$ \\
\noalign{\medskip}

Gauss/Diskl &$E_4$ (keV) & $2.74^{+0.01}_{-0.03}$ & $2.68\pm0.03$ & $2.67\pm0.03$ & $2.67^{+0.01}_{-0.02}$& $2.67\pm0.03$\\
Gauss/Diskl &$\sigma_4$ (keV) & (0.0) & ... & ... & ...& ...\\
Gauss/Diskl &$N_4$  & $1.0^{+0.2}_{-0.1}$ & $2.0\pm0.4$ & $2.3\pm0.4$ &$1.2\pm0.4$&$2.2\pm0.3$ \\
Gauss/Diskl &${EW}_4$ (eV)&$3.8\pm0.2$  & $7.5\pm0.4$ & $8.4\pm1.1$& $4.2\pm0.9$& $8.0\pm1.3$ \\
\noalign{\medskip}

Gauss/Diskl &$E_5$ (keV) & $3.30\pm0.03$ & $3.29\pm0.02$ & $3.27\pm0.04$& $3.28\pm0.03$  & $3.29\pm0.03$ \\
Gauss/Diskl &$\sigma_5$ (keV) & $0.13^{+0.04}_{-0.02}$ & ... & ... & ...& ... \\
Gauss/Diskl &$N_5$  & $2.5^{+0.6}_{-0.7}$ & $2.1\pm0.3$ & $2.1\pm0.3$& $1.7^{+0.2}_{-0.5}$& $1.8^{+0.3}_{-0.5}$\\
Gauss/Diskl &${EW}_5$ (eV) & $11.5\pm0.1$  & $9.2\pm0.1$ & $9.5\pm1.2$ & $7.2\pm1.1$& $8.8\pm1.1$ \\
\noalign{\medskip}

Gauss/Diskl &$E_6$ (keV) & $3.94^{+0.05}_{-0.06}$ & $3.96\pm0.02$ & $3.96\pm0.05$ & $4.01\pm0.05$& $3.96\pm0.05$ \\
Gauss/Diskl &$\sigma_6$ (keV) & $0.26^{+0.10}_{-0.07}$ &... & ... & ...& ...\\
Gauss/Diskl &$N_6$  &$2.8^{+1.8}_{-0.9}$ & $1.6\pm0.3$& $2.21\pm0.05$& $1.2\pm0.4$& $1.5^{+0.1}_{-0.3}$ \\
Gauss/Diskl &${EW}_6$ (eV)& $15.4\pm0.2$  & $8.5\pm0.1$ & $8.3\pm0.9$ & $6.3\pm1.9$& $8.1\pm1.2$ \\
\noalign{\medskip}

\noalign{\medskip}

  &$Flux$  & $9.34\pm0.01$ & $9.23\pm0.03$ & $26\pm3$ & $28\pm3$  &  $26\pm3$\\
\noalign{\medskip}

  &$\chi^2$ (d.o.f.) &  1.457 (106) & 1.338 (106) & 1.152 (1083)& 1.173 (1083)& 1.1487 (1081)\\
  &p$_{\mbox{null}}$ & $1.5\times10^{-3}$ & $1.1\times10^{-2}$ &$3.6\times10^{-4}$ & $6.1\times10^{-5}$& $4.6\times10^{-4}$\\
\noalign{\medskip}

\hline           
\label{tab:epn}       
\end{tabular}

\tablefoot{Fluxes are unabsorbed and expressed in units of
  $10^{-10}$~erg~cm$^{-2}$~s$^{-1}$. $F_{nthComp}$ is the flux in the Comptonization 
  component expressed as a fraction 
  of total flux.  For the fits of the EPIC-pn  spectrum alone (second and third column) 
  the fluxes are evaluated in the 0.5--10 keV energy band, while these are calculated in the
  0.5--100 keV range for the broadband spectrum (from the fourth to the seventh 
  column). The normalization of the lines are expressed in units of
  $10^{-4}$~ph~cm$^{-2}$~s$^{-1}$.}
\end{table*}

\normalsize

The broadness of the $6.75$ keV Fe XXV line suggests reflection of
hard X-rays off the inner parts of the accretion disk as a plausible
origin. We then replaced the Gaussian profile with a relativistic
broadened \textsc{diskline} profile \citep{Fabian.etal:89}. The three
emission lines found between 2.4 and 4 keV have high ionization states
and probably originate from the same region. We then modeled them
with relativistic broadened emission features as well, keeping the
disk emissivity index, $\beta_{irr}$, and the geometrical disk
parameters (the inner and outer disk radii, $R_{in}$ and $R_{out}$,
and inclination, $i$) tied to the values obtained for the Fe XXV
line. As the spectral fit was insensitive to the outer disc radius
parameter, we left it frozen to its maximum value allowed
($10^{7}$~R$_g$, where $R_g=GM/c^2$ is the NS gravitational
radius). Modeling of the neutral (or weakly ionized) narrow Fe lines
at $\simeq6.5$ and $7.1$ keV with a Gaussian profile was maintained.
We found that the energy of the lines were all consistent within the
uncertainties with those previously determined with Model I. The
parameters of the relativistic lines indicate a disk extending down
to $R_{in}=20_{-6}^{+4}$~R$_g$ with an inclination of
$i=(37\pm2)^{\circ}$ and an emissivity index of
$\beta=-2.44^{+0.04}_{-0.06}$ (see column dubbed Model II of
Tab.~\ref{tab:epn} for the whole list of parameters). Modeling of the
spectrum with these broad emission lines decreased the fit $\chi^2$ to
141.8, for 106 degrees of freedom, which translates
into a probability of $p_{null}=10^{-2}$ of obtaining a value of the
fit $\chi^2$ as large or larger if the data are drawn from such a
spectral model. Figure~\ref{fig:epn} shows the observed spectrum, the
residuals with and without the inclusion of the emission lines. The
model parameters are listed in the third column of
Table~\ref{tab:epn}.

\subsection{The 0.35--180 keV XMM-Newton/INTEGRAL broadband spectrum}
\label{sec:broad}

In order to study the broadband spectrum of {\exo} we fitted
simultaneously the spectra observed by the two RGS cameras (0.35--2.0
keV) and the EPIC-pn (2.4--10 keV) on-board {\xmm}, together with the
spectra observed by the two JEM-X cameras (5--25 keV) and ISGRI
(20-180 keV) on board {\igr} during the satellite revolution 1521,
which partly overlapped with the {\xmm} pointing. We initially
considered Model II, in which the continuum was modeled by the sum of
a Comptonized and a thermal component, the lines with energies
compatible with ionized species were described by a relativistic
broadened disk emission lines, and the K-$\alpha$ and K-$\beta$
lines of neutral (or weakly) ionized iron were modeled by a Gaussian
profile. For the broadband spectrum we decided to fix the energy 
of the K$\beta$ iron line to its rest-frame energy of $7.06$ keV to make 
the fit more stable.
The inclusion of the RGS spectra at low energies yielded a
measure of the equivalent hydrogen column density
$N_H=(2.02\pm0.04)\times10^{22}$~cm$^{-2}$. At the high energy end of
the spectrum, the ISGRI spectrum constrained the electron temperature
of the Comptonizing electron population to
$kT_e=37^{+7}_{-5}$~keV. The other parameters describing the continuum
and the lines were found to be compatible with those obtained from the
modeling of the EPIC-pn spectra alone. The model parameters are
listed in the fourth column of Table~\ref{tab:epn}, dubbed Model II*.

In order to entertain the hypothesis that the broad emission lines are
due to reflection of the primary Comptonized spectrum onto the inner
accretion disk, we replaced the Fe XXV broad emission line described
as {\textsc disklines} in Model II* with a self-consistent model
describing the reflection off an ionized accretion disk.
We convolved the Comptonized component describing the main source of
hard photons, \textsc{nthComp}, with the disk reflection model
\textsc{rfxconv} \citep{2011MNRAS.416..311K}.

\begin{figure*}
\centering

\includegraphics[scale=0.35, angle=-90]{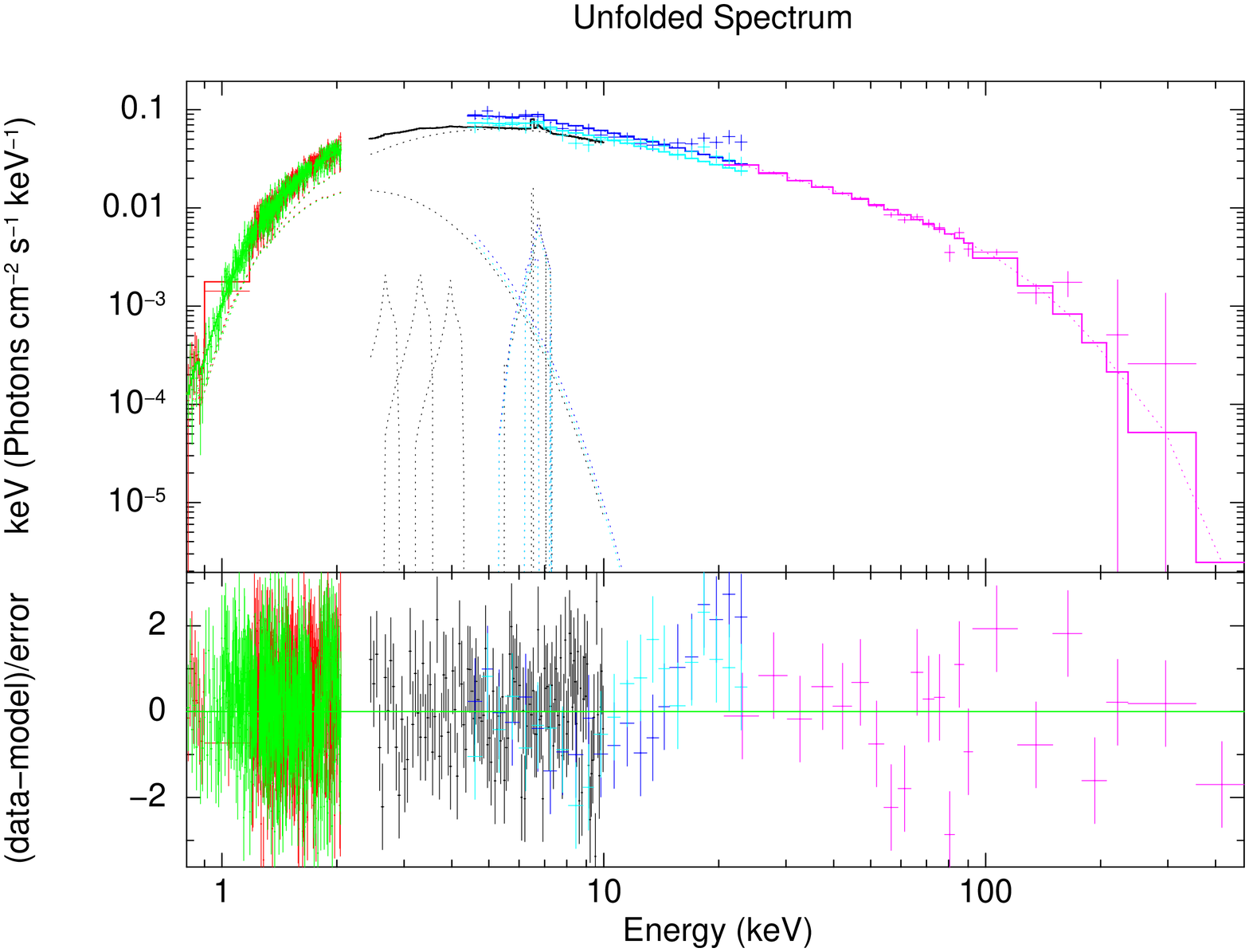}
 
\includegraphics[scale=0.35, angle=-90]{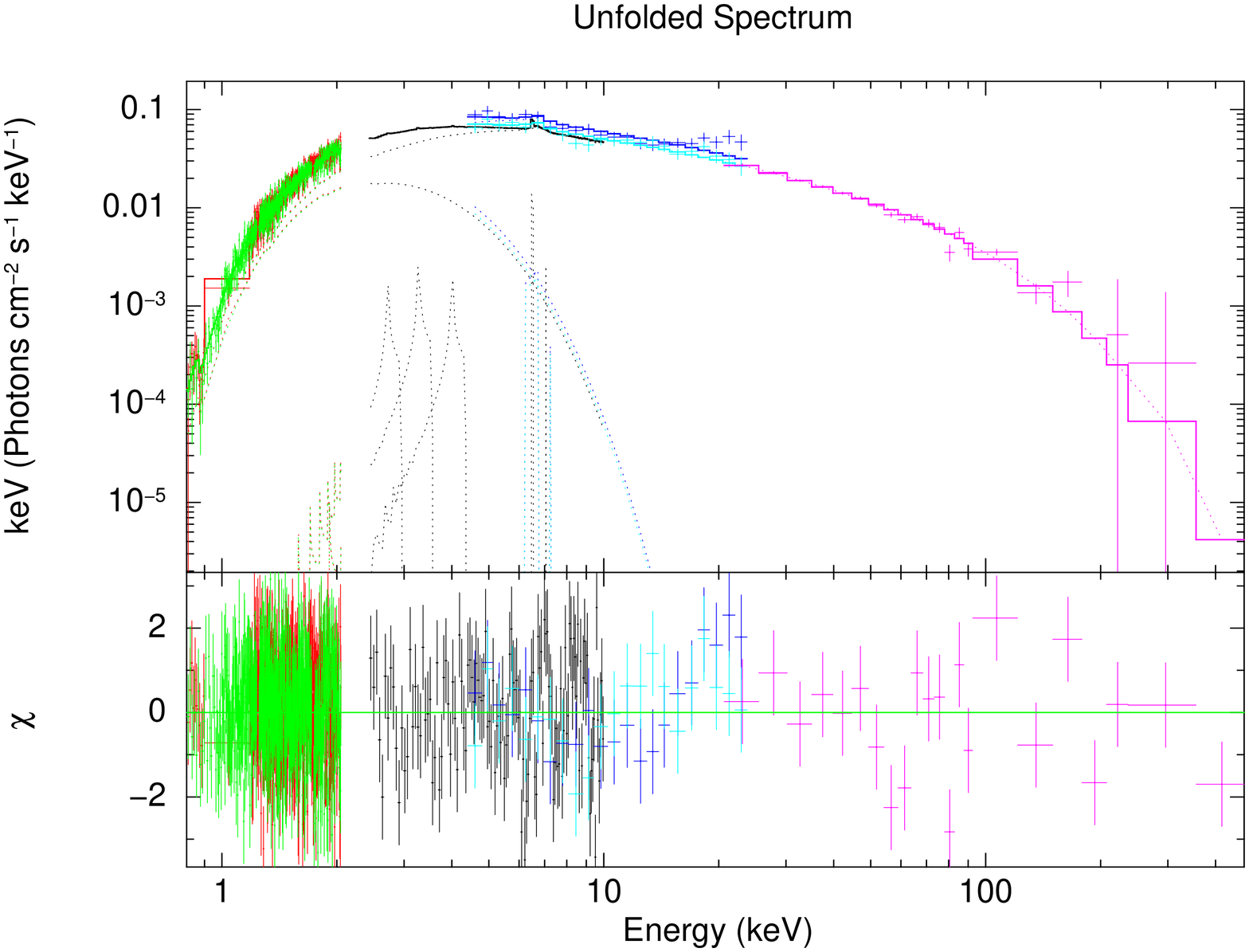}

\includegraphics[scale=0.35, angle=-90]{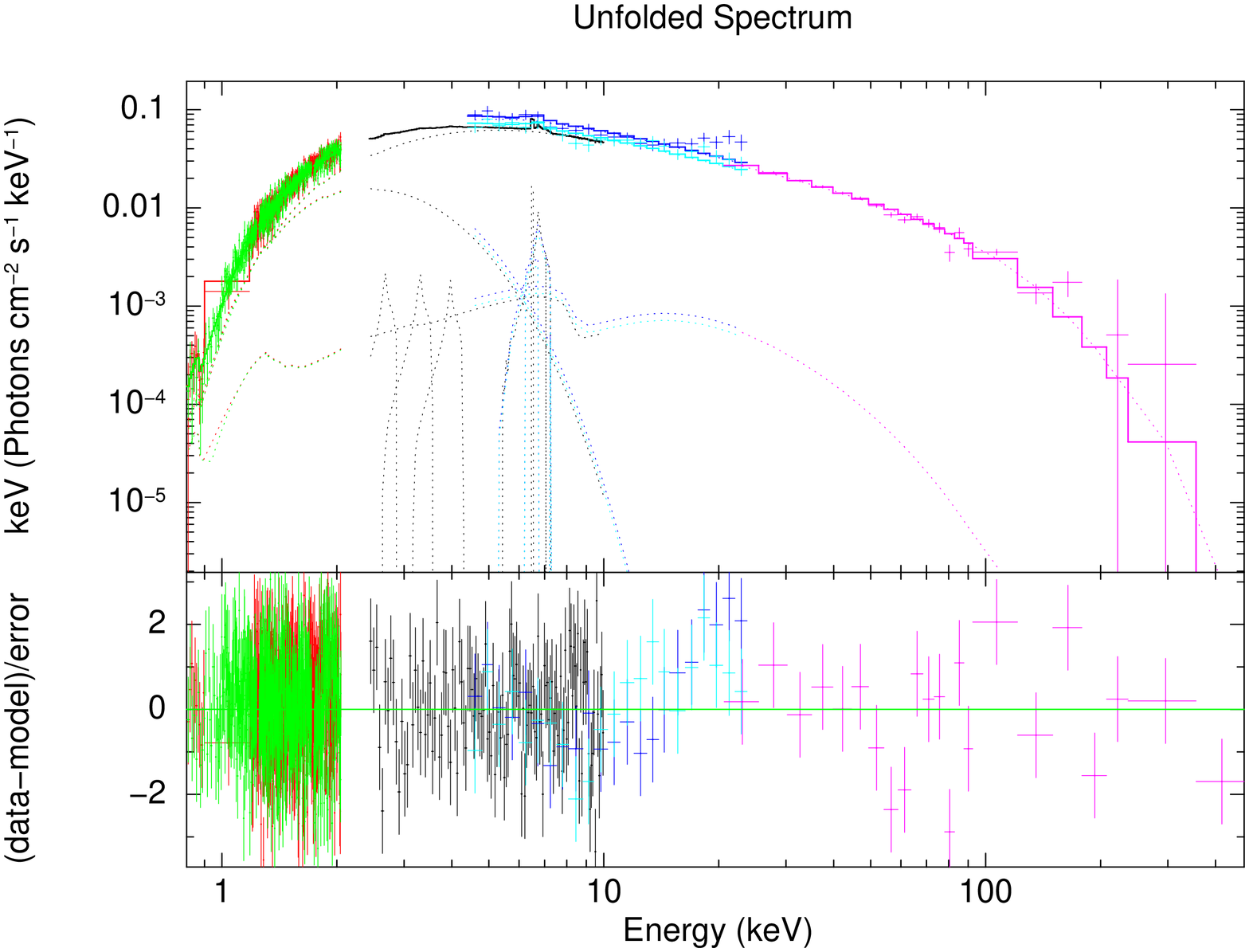}

\caption{Broadband spectra, models and residuals in units of $\sigma$ with respect to Model II*, Model III and Model IV are plotted in the top panel, middle panel and lower panel, respectively. RGS1 (red), RGS2 (green), EPIC-pn (black), JEMX1 (blue), JEMX2 (cyan) and ISGRI (magenta) spectra. }

    \label{fig:rfxconv}

\end{figure*}

We further convolved the
\textsc{rfxconv} component with a relativistic kernel
(\textsc{rdblur}) to take into account relativistic distortion of the
reflection component due to a rotating disc.  Because the
\textsc{rfxconv} model does not include Ar and Ca transitions
and does not give a good modeling of the S line, leaving clear residuals at $\sim 2.7$ keV,,
we included three \textsc{diskline} components for them, linking the
parameters of the \textsc{rdblur} component to the corresponding
smearing parameters of the \textsc{disklines} , according to the 
hypothesis that all these lines originate from the same disk region \citep[see,
  e.g.][]{disalvo2009,egron2013,disalvo2015}.
The best fit with this model (dubbed Model III, see sixth column of
Table~\ref{tab:epn}) was slightly worse ($\chi^{2}$
/dof =1271/1083) than for Model II* ($\chi^{2}$
/dof =1248/1083).  According to the reflection model, the solid angle
($\Omega_r/2\pi$) subtended by the reflector as seen from the
illuminating source was 0.22$\pm$ 0.04.  The logarithm of the
ionization parameter of the disc was $\simeq$ 2.7, which could well
explain the ionization state of the Fe XXV, S XVI, Ar XVIII and Ca XX
(or XIX) emission lines observed in the spectrum. The inclination
angle of the system was found to be consistent with 37$^{\circ}$.  The
broadband continuum and the line parameters were not significantly
changed by the introduction of the reflection model. The six instruments 
spectrum, Model III and residuals are plotted in Fig.~\ref{fig:rfxconv}. 
This figure shows  a clear trend of residuals between 10 and 
a 30 keV (consistent with a possible reflection hump), which our
reflection models cannot describe well.

To test independently the significance of the Compton hump and absorption 
edges, constituting the continuum of the reflection component, 
we also tried a different   reflection model, namely \texttt{pexriv}
\citep{Magdziarz.etal:95}, which describes an exponentially cut off
power law spectrum reflected from ionized material. We fixed the disk
temperature to the default value $ 10^{6}$ K, and the value on
reflection fraction to 0.22, that is the best value found in Model
III. We also tied parameters describing the irradiating power-law
(photon index and energy cut-off) to those indicated by the
\texttt{nthComp} component. As the iron emission is not included in
the \texttt{pexriv} model, we added a \texttt{diskline} centered at
6.75 keV.  The results of the fit are reported in the sixth column
of Table~\ref{tab:epn}, labeled 'Model IV'. The parameters describing
the irradiating continuum and the reflection component are compatible
with those obtained with \textsc{rfxconv}, and the fit $\chi^2$
slightly improved with respect to Model III ($\Delta\chi^2=28.6$ for
two degrees of freedom less), while is compatible with the results
obtained with Model II*.

Since the width of the Gaussian lines describing the K$\alpha$
and K$\beta$ lines from neutral or mildly ionized iron were always compatible
with 0, we fixed the width of these lines to be null in all our fits.
This is going to be an acceptable assumption because the upper limits 
calculated for the width of the iron K$\alpha$ line at 6.5 keV are 
$\sigma <  0.022$  keV and $\sigma  <  0.040$  keV,
for Model III and Model IV , respectively.


\section{Temporal analysis}
\label{sec:timing}

\begin{figure}
 \resizebox{\hsize}{!} 
{\includegraphics{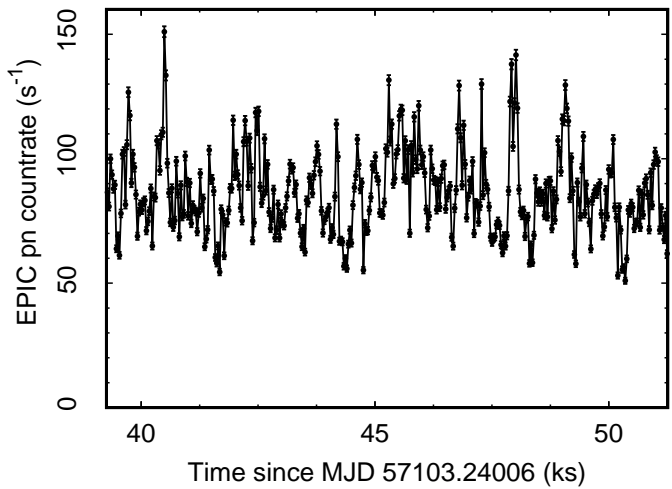}}
\vskip 1cm
\caption{Snapshot of the 0.5-10 keV {\it persistent} light curve
  observed by the EPIC-pn on-board XMM-Newton. Counts were binned in
  32 s-long intervals.}
\label{fig:perslc}
\end{figure}

\begin{figure}
 \resizebox{\hsize}{!} 
{\includegraphics{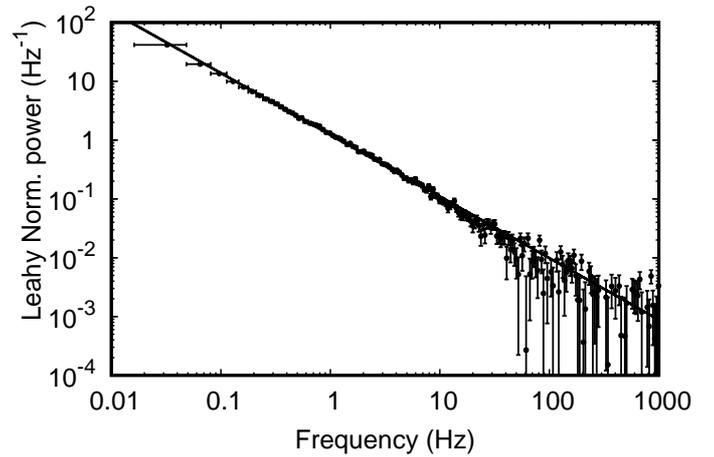}}
\caption{Leahy normalized power density spectrum evaluated averaging
  spectra computed over 8~s-long intervals of the EPIC-pn observation,
  and re-binning the resulting spectrum as a geometrical series with
  ratio equal to 1.04. A white noise level equal to 1.99(1) has been
  subtracted. The solid line represents a power law,
  $P(\nu)=\nu^{-\alpha}$ with index $\alpha=1.05$.}
\label{fig:pds}
\end{figure}

The {\it persistent} (i.e, non-bursting) emission observed during the
XMM-Newton EPIC-pn observation was highly variable, with a sample
fractional rms  amplitude of 0.33.  A portion of the {\it
  persistent} light curve is shown in Fig.~\ref{fig:perslc} for
illustrative purposes. To study the power spectrum of the aperiodic
variability we performed a fast Fourier transform of 32-s long
intervals of the 0.5--10 keV EPIC-pn time series with 59~$\mu$s time
resolution (corresponding to a Nyquist maximum frequency of
8468~Hz). We averaged the spectra obtained in the various intervals,
re-binning the resulting spectrum as a geometrical series with a ratio
of 1.04. The Leahy normalized and white noise subtracted average power
spectrum is plotted in Fig.~\ref{fig:pds}. The spectrum is dominated
by a flicker noise component described by a power law,
$P(\nu)\propto\nu^{-\alpha}$, with $\alpha=1.05(1)$, slightly
flattening towards low frequencies. In order to search for kHz quasi
periodic oscillations already observed from the source at a frequency
ranging from 690 to 715~Hz \citep{mukherjee2011,barret2012}, we
produced a power density spectrum over 4 s-long intervals to have a
frequency resolution of 0.25~Hz, and averaged the spectra extracted
every 40 consecutive intervals. No oscillation was found within a
3-$\sigma$ confidence level upper limit of 1.5\% on the rms variation.

In order to search for a coherent signal in the light curve obtained
by the EPIC-pn, we first reported the observed photons to the Solar
system barycenter, using the position RA=17$^h$~48$^m$~05.236,
DEC=-24$^{\circ}$~46'~47.38" reported by \citet{Heinke.etal:06b} with
an uncertainty of 0.02" at 1-$\sigma$ confidence level). We performed
a power density spectrum on the whole $t_{pds}=77.5$~ks exposure,
re-binning the time series to a resolution equal to eight times the
minimum ($t_{res}=2.3\times10^{-4}$~s, giving a maximum frequency of
$\nu_{Ny}=2117$~Hz). After taking into account the number of
frequencies searched, $N_f\simeq\nu_{Ny}\,t_{pds}=1.64\times10^8$, 
we could not find any significant signal with an
upper limit at 3-$\sigma$ confidence level of 0.5\% on the amplitude
of a sinusoidal signal, evaluated following \citet{1994ApJ...435..362V}.

The orbital period of {\exo} is currently unknown. On the
spectral properties, \citet{Heinke.etal:06b} suggested it might be
hosted in an ultra-compact binary ($P_{orb}<<1$~d). Based on empirical
relation between the V magnitude of the optical counterpart, the X-ray
luminosity and the orbital period, \citet{ferraro2015} estimated a
likely range for the orbital period between 0.1 and 1.3~d. As the
orbital period is likely of the same order of the length of the exposure 
of the observation considered, or shorter, the orbital motion will induce
shifts of the frequency of a coherent signal that hamper any
periodicity search. We then performed a search on shorter time intervals,
with a length ranging from 124 to 5500~s. The data acquired during 
type-I X-ray bursts was discarded. No signal was detected at a confidence
level of 3-$\sigma$, with an upper limit ranging between 14\% and 2\%, with 
the latter limit relative to the longer integration time.

In order to improve the sensitivity to signals affected by the unknown
binary orbital motion, we applied the quadratic coherence recovery
technique described by \citet{1991ApJ...379..295W} and
\citet{1994ApJ...435..362V}.  We divided the entire light curve in
time intervals of length equal to $\Delta t=495$~s. In each of the
intervals the time of arrival of X-ray photons $t_{arr}$ were
corrected using the relation $t'=\alpha t_{arr}^2$; the parameter
$\alpha$ was varied in steps equal to $\delta\alpha=(2\nu_{Ny}\Delta
t^2)^{-1}=9.6\times10^{-10}$~s$^{-1}$ to cover a range between
$\alpha_{max}=1.7\times10^{-8}$~s$^{-1}$ and
$\alpha_{min}=-\alpha_{max}$. The width of the range is determined by
a guess on the orbital parameters of the system that would be optimal
for an orbital period of $12$~h, a donor star mass of
$M_2=0.3$~M$_{\odot}$, a NS spin period of $P=3$~ms, and a donor to NS
mass ratio of $q=0.2$ (see Eq.~14 of \citealt{1991ApJ...379..295W}). 
This method confirmed the lack of any significant periodic signal, 
with an 7\% on the sinusoidal
amplitude. We also considered a shorter time interval of
$\Delta~t=247$~s, and still obtained no detection within a 3$\sigma$
c.l. upper limit of 10.5\%.

We also searched for burst oscillations in the seven events observed
during the {\xmm} exposure. To this aim, we produced power density
spectra over intervals of variable length, ranging from 2 to 8~s, and
time resolution equal to that used above
($t_{res}=2.3\times10^{-4}$~s). No significant signal was detected in
either of the bursts, with 3$\sigma$ c.l. upper limit on the signal
amplitude of the order of $\simeq20$ and $\simeq10\%$ for the
shorter and longer integration times used, respectively.

\section{Type I X-ray bursts}
\label{sec:burst}
\begin{figure}
 \resizebox{\hsize}{!} 
{\includegraphics{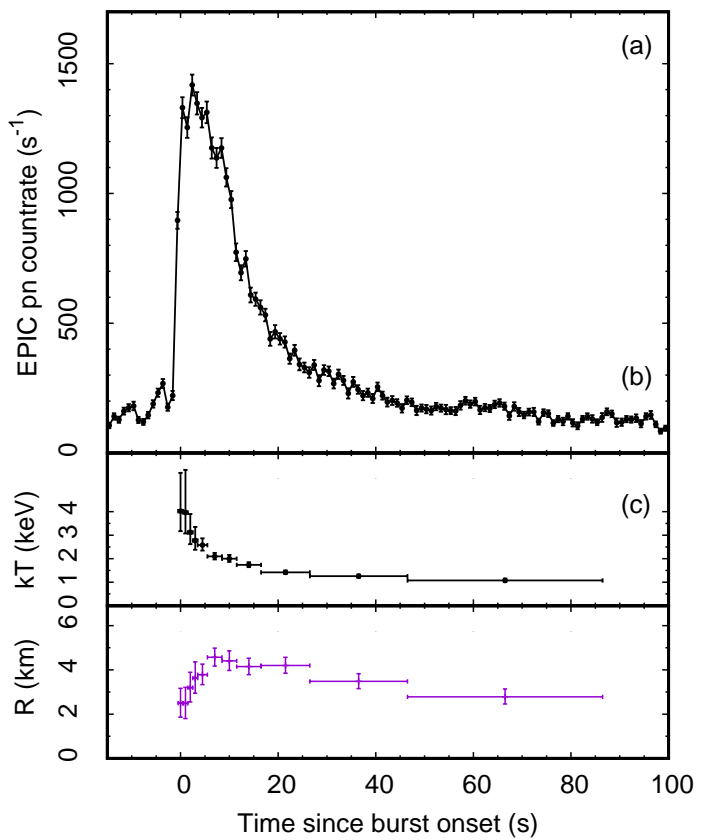}}
\caption{0.5-10 keV light curve of the second burst observed by the
  EPIC-pn, which begun on $T_2=57103.41516$~MJD (top panel). The
  central and bottom panels show the temperature and apparent radius
  of the black body used to model the time-resolved spectra,
  respectively. The radius is evaluated for a distance of 5.5 kpc.
  Errors are reported with a 90\% confidence.}
    \label{fig:burst}
\end{figure}

Seven bursts took place during the {\xmm} observation, with a
recurrence time varying between $t_{rec}=2.5$ and 4 hours (see
Table~\ref{tab:bursts}). The bursts attained a peak 0.5--10 keV
EPIC-pn count rate ranging from 1100 to 1500 counts/s (see top panel
of Fig.~\ref{fig:burst} where we plot the light curve of 
the second burst seen during the {\xmm} exposure). Such values exceed the
EPIC-pn telemetry limit ($\approx$450 counts/s; ), and data overflows
occurred close to the burst maximum. The burst rise takes place in
less than $\approx5$~s, while the decay could be approximately modeled
with an exponential function with an e-folding time scale ranging
between 10 and 23~s.
In order to analyze the evolution of the spectral shape during 
the bursts, we extracted spectra over time intervals of length ranging 
from 1 to 100 s depending on the count rate.
In order to minimize the
effect of pile up, which becomes important when the count rate
increases above a few hundreds of counts per second, we removed the
two brightest columns of the EPIC-pn chip (RAWX=36-37). Background was
extracted considering the {\it persistent emission} observed between
600 and 100~s before the burst onset. The resulting spectra were
modeled with an absorbed black-body, fixing the absorption column to
the value found in the analysis of the {\it persistent} emission
($N_H=2\times10^{22}$~cm$^{-2}$). The evolution of the temperature and
apparent radius observed during the second burst, the one with the
highest peak flux seen in the {\xmm} observation, are plotted in the
middle and bottom panels of Fig.~\ref{fig:burst}, respectively.  The
temperature attained a maximum value of $\approx{3.5}~$keV and then
decreased steadily, confirming the thermonuclear nature of the
bursts. The estimated apparent extension of the black-body emission 
remained always much lower than any reasonable value expected for the  
radius of a standard neutron star ($\gtrsim$8-13~km). The maximum flux 
attained a value of $3.8(7) \times 10^{-8}$ {\fluxcgs} (see Table~\ref{tab:bursts}), 
which translate into a luminosity of
$1.4(2)\times10^{38}\,d_{5.5}^2$~{\lumcgs}. This value is both lower
than the Eddington limit for a NS and cosmic abundance
($1.76\times10^{38}\,(M/1.4\,M_{\odot})$~{\lumcgs}) and the luminosity
attained during the two bursts characterized by photospheric radius
expansion reported by \citet[][$L_{\rm pre}\simeq
  2.2\times10^{38}\,d_{5.5}^2$~{\lumcgs}]{galloway2008}.  Similar properties were
observed also in the other bursts and we concluded that photospheric
radius expansion did not occur in any of the bursts observed by {\xmm}.
In addition, bursts characterized by photospheric radius expansion often 
show a distinctive spectral evolution after the rise, characterized by a dip of 
the black-body temperature occurring when the radius attains its maximum value, 
while the flux stays at an approximately constant level (see, e.g., Fig. 2 of 
Galloway et al. 2008). Although the minimum time resolution of our spectral 
analysis (1s) is limited to by the available photon statistics, such a 
variability pattern does not seem to be present in the observed evolution 
of the burst parameters (see Fig.~7). Combined to the relatively low X-ray 
luminosity attained at the peak of the bursts, we then conclude that it is 
unlikely that photospheric radius expansion occurred in any of the bursts 
observed by {\xmm}.

Table~\ref{tab:bursts} lists the energetics of the seven bursts
observed by {\xmm}.  The persistent flux was evaluated by fitting the
spectrum observed from 500~s after the previous burst onset, and 50~s
before the actual burst start time, using Model I (see
Table~\ref{tab:epn}). We measured the fluence $\mathcal{F}$ by summing
the fluxes observed in the different intervals over the duration of
each burst. We also evaluated the burst timescale as the ratio
$\tau=\mathcal{F}/F_{\rm peak}$ \citep{1988MNRAS.233..437V}. The
rightmost column of table~\ref{tab:bursts} displays the parameter
$\alpha$, defined as the ratio between the persistent integrated
flux and the burst fluence \citep[$\alpha=c_{\rm bol}F_{\rm
    pers}t_{rec}/\mathcal{F}$; see, e.g.,][]{galloway2008}, where
$c_{\rm bol}$ is a bolometric correction factor that we estimated from
the ratio between the flux observed in the 0.5--100 keV and the
0.5--10 keV  band with Model II* and II, respectively (see
Table~\ref{tab:epn}), $c_{\rm bol}=2.8\pm0.3$. We evaluated values of
$\alpha$ ranging between 50 and 110, with an average $<\alpha>=82$.

\begin{table*}[!h]
  \caption{Properties of the type-I X-ray bursts observed by XMM-Newton. \label{tab:bursts}}

\centering          

\begin{tabular}{lccccccc}     
  \hline\hline

  No. & Start time (MJD) & $t_{rec}$~(s) & $F_{\rm pers}$ & $F_{\rm peak}$ & $\mathcal{F}$ & $\tau$ (s)  & $\alpha$\\

  \hline
  I   & 57103.26624 & ...   & 0.99(2)  & 17(2) & 38(3) & $22.7\pm3.4$  &   \\ 
  II  & 57103.41516 & 12866 & 0.955(7) & 38(7) & 40(6) & $10.5\pm2.6$  &  $86\pm16$ \\ 
  III & 57103.56912 & 13303 & 0.924(6) & 18(2) & 31(3) & $16.5\pm3.0$  &  $111\pm12$ \\ 
  IV  & 57103.67557 & 9197  & 0.91(1)  & 21(3) & 42(5) & $20.4\pm3.8$  &  $56\pm6$ \\ 
  V   & 57103.84017 & 14221 & 0.868(4) & 29(4) & 44(5) & $18.8\pm2.8$  &  $79\pm9$ \\ 
  VI  & 57103.96830 & 11071 & 0.929(5) & 24(3) & 38(4) & $16.2\pm2.8$  &  $76\pm11$ \\ 
  VII & 57104.10384 & 11710 & 0.922(4) & 22(3) & 37(4) & $17.1\pm3.1$  &  $82\pm12$ \\ 
  
\hline                  
\end{tabular}

\tablefoot{The 0.5--10 keV persistent flux $F_{pers}$ and the burst
  peak flux $F_{peak}$ are unabsorbed and expressed in units of
  $10^{-9}$~{\fluxcgs}. The bolometric fluence $\mathcal{F}$ is
  unabsorbed and given in units of $10^{-8}$~erg~cm$^{-2}$. The burst
  decay timescale was evaluated as $\tau=\mathcal{F}/F_{pers}$, the
  parameter $\alpha$ as $c_{bol}F_{pers}t_{rec}/\mathcal{F}$ with
  $c_{bol}=2.8\pm0.3$. }
\end{table*}

\section{Discussion}

We analyzed quasi-simultaneous  {\xmm} and {\igr} observations of the transient LMXB
{\exo} in the massive globular cluster Terzan 5, carried out when the
source was in the hard state, just after it went into outburst in
2015, with the aim to characterize its broad-band spectrum and its
temporal variability properties.
We also made use of all additionally available INTEGRAL data collected 
during the outburst of the source in 2015
to spectroscopically confirm its hard-to-soft state transition occurred around 57131~MJD. 
This transition was firstly noticed
by \citet{tetarenko2016} using the source lightcurves extracted from Swift/BAT, Swift/XRT, and MAXI.
\subsection{The combined XMM-Newton and INTEGRAL spectrum}

We modeled the spectrum observed simultaneously by {\xmm} and {\igr}
to study the X-ray emission from the source in the energy range
0.8--100 keV.  We estimated an unabsorbed total luminosity ($0.5-100$ keV
energy range) of $\approx 1 \times
10^{37}\,d_{5.5}^2$~{\lumcgs}. The continuum was well described by a
two-component model, corrected by the low-energy effects of
interstellar absorption. The best-fit value of the equivalent hydrogen
column density, $N_H$, is $(2.02\pm0.05) \times 10^{22}$~cm$^{-2}$,
slightly lower than the estimate of interstellar absorption towards
Terzan 5 given by \citet[][]{2014ApJ...780..127B},
$N_H=(2.6\pm0.1)\times 10^{22}$~cm$^{-2}$.  The two-component
continuum model consist of a quite hard Comptonization component,
described by the \texttt{nthComp} model, with electron temperature
$kT_e \sim 40$ keV, photon index $\Gamma \simeq 1.8-2$ and seed-photon
temperature of about 1.3 keV, and of a soft thermal component
described by a black-body with temperature $kT \sim 0.6-0.7$ keV. The
Comptonization component contributed to more than 90 per cent of the
flux observed during the observations considered, clearly indicated
that the source stayed in the hard state.

Assuming a spherical geometry for both the black-body and the
seed-photon emitting regions, and ignoring any correction factor
due to color temperature corrections or boundary conditions, we found a
radius of the black-body emitting region of about $R_{bb}=3.5-5$ km and a
radius of the seed-photon emitting region of about $R_w=2-3$ km. 
Given these modest extensions, it is likely that the surfaces of seed 
photons are related to hot spots onto the neutron star surface.
The latter was calculated using the relation reported by
\citet{1999A&A...345..100I}, assuming an optical depth of the
Comptonization region, $\tau=2.2\pm0.3$, evaluated using the relation
between the optical depth, the temperature of the Comptonizing
electrons and the asymptotic power-law index given by
\citet{1987ApJ...319..643L}.

A similar spectral shape was found during the 2000 outburst of {\exo}
observed by {\it Chandra} and {\it RXTE} \citep{Heinke.etal:03a}.  In
that case the continuum model consisted of a multicolor disk
black-body, characterized by an inner temperature of $kT = 0.6 - 1.2$
keV and an inner disk radius of $r_{in}/d_{10}(\cos{i})^{0.5} = 4.3 -
9.2$ km, and a Comptonization component, described by the
\texttt{comptt} model, characterized by a seed photon temperature of
$kT_{0} = 1.2 - 1.7$ keV and radius $R_{W}$ = 3.1 - 6.7 km, an
electron temperature of $kT_{e} = 9.8 - 10.7$ keV, and an optical
depth $\tau=8$. The Comptonization spectrum was softer during the
Chandra/RXTE observations than during the XMM-Newton/INTEGRAL
observation analyzed here, and the 0.1-100 keV luminosity was $L_X
\approx 6.6 \times 10^{37}$ erg/sec, higher by about a factor 6 than
during our observation. Such a softening of the Comptonization spectrum
with increasing luminosity is in agreement with the results presented 
by \citet{tetarenko2016}
for the 2015 outburst using Swift/XRT data (see their Table~1) and our 
findings in Sect.~\ref{sec:integral} by using the INTEGRAL monitoring data.

Thanks to the large effective area and the moderately-good
energy-resolution of the EPIC-pn, we could detect several emission
features in the spectrum of EXO~1745-248. Most of the emission
features are broad and identified with K$\alpha$ transitions of highly
ionized elements. These are the $2.6-2.7$ keV line identified as S XVI
transition (H-like, expected rest frame energy 2.62 keV), the 3.3 keV
line identified as Ar XVIII transition (H-like, expected rest-frame
energy 3.32 keV), the $3.96-4.1$ keV line identified as Ca XIX or Ca
XX transition (He or H-like, expected rest-frame energy $3.9$ and
$4.1$ keV, respectively), and the 6.75 keV line identified as Fe XXV
(He-like) transition (expected rest-frame energy 6.7 keV). The Gaussian
width of the Fe XXV line we observed from {\exo}, $\sigma_1 =
0.24^{+0.03}_{-0.02}$ keV, is compatible with the width of the Fe line
detected during the 2000 outburst \citep{Heinke.etal:03a}. The widths
of the low energy lines are compatible with being about half the width
of the iron line, in agreement with the expectations from Doppler or
thermal Compton broadening, for which the width is proportional to the
energy. Therefore all these lines are probably produced in the same
emitting region, characterized by similar velocity dispersion or
temperature (i.e., the accretion disk).

The fitting of the iron line appears, however, much more complex and
puzzling than usual. At least two components are needed to fit the
iron emission feature because of highly significant residuals still
present after the inclusion in the model of a broad Gaussian. We
fitted these residuals using another Gaussian centered at $\sim 6.5$
keV (therefore to be ascribed to neutral or mildly ionized iron) which
appears to be much narrower than the previous component (its width is
well below the energy resolution of the instrument and compatible with
0). Driven by a small residual still present at $\sim 7$ keV and by
the expectation that the 6.5-keV $K\alpha$ transition should be
accompanied by a 7.1-keV $K\beta$ transition, we also added to
the model a narrow Gaussian centered at $\sim 7.1$ keV, which we
identify with the K$\beta$ transition of neutral or mildly ionized
iron. Note that the flux ratio of the K$\beta$ transition to the
K$\alpha$ transition reaches its maximum of $0.15-0.17$ for Fe VIII,
while it drops to less than 0.1 for charge numbers higher than Fe X-XI
\citep[see][]{palmeri03}. This suggest that these components originate
from low-ionization iron (most probably Fe I-VIII) and come from a
different region, plausibly farther from the ionizing central engine,
with respect to the other broad and ionized emission lines.

In the hypothesis that the width of the broad lines is due to Doppler
and relativistic smearing in the inner accretion disk, we fitted these
lines in the EPIC-pn spectrum using relativistic broadened
disk-lines instead of Gaussian lines (see Model II and II* in
Table~\ref{tab:epn}).  We obtained a slight improvement of the fit.
According to this model we obtained the emissivity index
of the disk, $\propto r^{\beta}$ with $\beta \sim -2.4$, the inner
radius of the disk, $R_{in} \sim 14-24$ R$_g$, and the inclination
angle of the system, $\sim 37^\circ$. 
Although we have hints for the inclination angle to be relatively 
low (e.g., lack of intrinsic absorption, dip activity or eclipses), it is 
worth noting that values of the inclination angle of the system derived from 
spectral fitting of the reflection component may rely on the assumed geometry 
of the disk-corona system and therefore uncertainties on this parameter may
be underestimated.
Taking advantage from the broad-band coverage ensured by the almost
simultaneous XMM-Newton and INTEGRAL spectra, we also attempted to use
a self-consistent reflection model, which takes into account both the
discrete features (emission lines and absorption edges, as well as
Compton broadening of all these features) and the Compton scattered
continuum produced by the reflection of the primary Comptonized
spectrum off a cold accretion disk (Model III in Table~\ref{tab:epn}).
However, we could not obtain a statistically significant improvement
of the fit with respect to the disklines model. All
the parameters were similar to those obtained with the diskline
model.  The only change in the smearing parameters we get using the
reflection model instead of disklines is in the value of the inner
disk radius, which is now constrained to be $< 8.5$ R$_g$.  The
reflection component required a ionization parameter of $\log \xi \sim
2.7$, consistent the high ionization degree of the broad lines, and a
reflection fraction (that is the solid angle subtended by the
reflector as seen from the corona, $\Omega / 2 \pi$) of about 0.22.  A
non significant improvement in the description of the spectrum ($\Delta
\chi^2 \simeq -5$ for the addition of two parameters) was obtained
when using  {\texttt pexriv} to model the reflection continuum (Model
IV, see Table~\ref{tab:epn} with respect to best fit model (Model II* 
in Tab~\ref{tab:epn}).  
The observation analyzed here were then not sufficient to ascertain with 
statistical significance whether a reflection continuum is present in the 
spectrum.


The smearing parameters of the reflection component were similar
to what we find for other sources.  The emissivity index of the disk,
$\sim -2.5$, the inner radius of the disk, about 30 km or below 13 km,
according to the model used for the reflection component, as well as
the inclination with respect to the line of sight, $35-40^\circ$, are
similar to the corresponding values reported in literature for
many other sources \citep[see e.g.][and references
  therein]{disalvo2015}. For instance, in the case of atoll LMXB 4U
1705--44 the inner disk radius inferred from the reflection component
lay around $14-17$ R$_g$ both in the soft and in the hard state,
changing very little (if any) in the transition from one state to the
other \citep{disalvo2009,egron2013,disalvo2015}. In the case of 4U
1728--34, caught by {\xmm} in a low-luminosity (most probably hard)
state, the inner disk radius was constrained to be $14-50$ R$_g$
\citep{egron2011}.  Even in the case of accreting millisecond pulsars
(AMSPs), which are usually found in a hard state and for which we
expect that the inner disk is truncated by the magnetic field, inner
disk radii in the range $6-40$ R$_g$ were usually found \citep[see,
  e.g.][]{papitto2009,cackett2009,papitto2010,papitto2013,pintore2016,king2016}.
Also, the reflection fraction inferred from the {\texttt rfxconv}
model, $\Omega/2\pi \sim 0.22$, although somewhat smaller than what is
expected for a geometry with a spherical corona surrounded by the
accretion disk ($\Omega/2\pi \sim 0.3$), is in agreement with typical
values for these sources. Values of the reflection fraction below or
equal 0.3 were found in a number of cases
\citep[e.g.][]{disalvo2015,degenaar2015,pintore2015,pintore2016,ludlam2016,chiang2016}.
More puzzling is the high ionization parameter required from the broad
emission lines, $\log \xi \sim 2.7-2.8$, where $\xi = (L_X / (n_e
r^2)$ is the ionization parameter, $L_X$ is the bolometric luminosity
of the central source and $n_e$ and $r$ are the electron density in
the emitting region and the distance of the latter from the central
source, respectively. This high value of the ionization parameter is
quite usual in the soft state, while in the hard state a lower
ionization is usually required, $\log \xi < 2$. This was clearly evident in the
hard state of 4U 1705—44 \citep{disalvo2015}, although in that case
the luminosity was $\sim 6 \times 10^{36}$ ergs/s, about a factor 2
below the observed luminosity of EXO~1745-248 during the observations
analysed here.

Perhaps the most unusual feature of this source is the
simultaneous presence in its spectrum of a broad ionized iron line and
at least one narrow, neutral or mildly ionized iron line, both in
emission and clearly produced in different regions of the
system. Sometimes, in highly inclined sources, broad iron emission
lines were found together with highly ionized iron lines in
absorption, clearly indicating the presence of an out-flowing disk wind
\citep[see, e.g., the case of the bright atoll source GX 13+1;][and
  references therein]{pintore2014}.  In the case of 4U 1636—536,
\citet{pandel2008} tentatively fitted the very broad emission feature
present in the range $4-9$ keV with a combination of several K$\alpha$
lines from iron in different ionization states. In particular they
fitted the iron complex with two broad emission lines with centroid
energies fixed at 6.4 and 7 keV, respectively. However, to our
knowledge, there is no other source with a line complex modeled by
one broad and one (or two) narrow emission features, as the one showed by
EXO~1745-248. While a natural explanation for the broad, ionized
component is reflection in the inner rings of the accretion disk, the
narrow features probably originate from illumination an outer region
in which the motion of the emitting material is much slower, as well 
as the corresponding ionization parameter. 
Future observation with instruments with a higher spectral resolution will be
needed to finely deconvolve the line shape, and firmly assess the
origin of each component.

\subsection{Temporal variability}

The high effective area of the EPIC-pn on board {\xmm}, combined with
its $\mu$s temporal resolution, make it the best instrument currently
flying to detect coherent X-ray pulsations, and in particular those
with a period of few milliseconds expected from low magnetic field NS
in LMXBs. We performed a thorough search for periodicity in the
EPIC-pn time series observed from {\exo}, but found no significant
signal. The upper limits on the pulse amplitude obtained range from
$2$ to $15\%$ depending on the length of the intervals considered, the
choice of which is a function of the unknown orbital period, and on
the application of techniques to minimize the decrease of sensitivity
to pulsations due to the orbital motion. Such upper limits are of the
order, and sometimes lower than the amplitudes usually observed from
AMSPs \citep[see, e.g.,][]{2012arXiv1206.2727P}. Though not excluding
the possibility of low amplitude pulsations, the non detection of a
signal does not favor the possibility that {\exo} hosts an observable
accreting millisecond pulsar (AMSP). This is also hinted by the
significantly larger peak luminosity reached by {\exo} during its
outbursts ($\sim 7\times10^{37}$~{\lumcgs}) with respect to AMSPs
($\approx\mbox{few}\times10^{36}$~{\lumcgs}). Together with the long
outburst usually shown (t$\sim100$~d), such a large X-ray luminosity
suggests that the long term accretion rate of {\exo} is more than ten
times larger than in AMSPs. A larger mass accretion rates is though to
screen the NS magnetic field \citep{2001ApJ...557..958C}, possibly
explaining why ms pulsations are observed only from relatively faint
transient LMXBs.

At the moment of writing this paper, the orbital parameters of {\exo}
were not known. In agreement with \citet{galloway2008} and
\citet{tetarenko2016},
we have reported a relatively long time scale of the X-ray bursts’ decay, 
indicating presence of Hydrogen and hence providing evidence against an
ultra-compact nature of this system.  
Recently, \citet{ferraro2015} showed that the location
of the optical counterpart of {\exo} in the color-magnitude diagram of
Terzan 5 is close to the cluster turnoff, and is compatible with a
0.9~M$_\odot$ sub-giant branch star if it belongs to the low
metallicity population of Terzan 5. In such a case the mass transfer
would have started only recently. The orbital period would be
$\sim0.9$ days and the optimal integration time to perform a search
for periodicity $\sim920\,(P_s/3\mbox{ms})^{1/2}$~s, where $P_s$ is
the spin period of the putative pulsar (when not performing an
acceleration search; see Eq.~21 in \citet{1991ApJ...368..504J},
evaluated for a sinusoidal signal and an inclination of
37$^{\circ}$). The upper limit on the signal amplitude we obtained by
performing a signal search on time intervals of this length is 5\%.

A useful comparison can be made considering the only accreting pulsar
known in Terzan 5, IGR~J17480--2446, a NS spinning at a period of
90~ms, hosted in a binary system with an orbital period of 21.3~hr
\citep{2011A&A...526L...3P}. Its optical counterpart in quiescence
also lies close to the cluster turnoff
\citep{2012A&A...547A..28T}. The relatively long spin period of this
pulsar and its relatively large magnetic field compared to AMSP, let
\citet{2012ApJ...752...33P} to argue that the source started to
accrete and spin-up less than a few $10^{7}$~yr, and was therefore
caught in the initial phase of the mass transfer process that could
possibly accelerate it to a spin period of few milliseconds. When the
IGR~J17480--2446 was found in a hard state, X-ray pulsations were
observed at an amplitude of 27 per cent, decreasing to a few per cent
after the source spectrum became softer and cut-off at few keV
\citep{2012MNRAS.423.1178P}. The upper limit on pulsations obtained
assuming for {\exo} similar parameters than IGR~J17480--244 is $2\%$,
of the order of the amplitude of the weaker pulsations observed from
IGR~J17480--244.

On the other hand, if the companion star belongs to the metal-rich
population of Terzan 5, it would be located in the color-magnitude
diagram at a position where companions to redback millisecond pulsars
are found \citep{ferraro2015}. In such a case a spin period of few
millisecond would be expected for the NS, and upper limits ranging
from 5 to 15\% on the pulse amplitude would be deduced from the
analysis presented here, depending on the orbital period. For
comparison, the redback transitional ms pulsar IGR J18245--2452 in the
globular cluster M28 showed pulsations with amplitude as high as 18\%,
that were easily detected in an {\xmm} observation of similar length
than the one presented here
\citep{2013Natur.501..517P,2014A&A...567A..77F}. This further suggests
that {\exo} is unlikely an observable accreting pulsar, unless its
pulsations are weak with respect to similar systems and/or it belongs
to a very compact binary system. Neither a search for burst
oscillations yielded to a detection, with an upper limit of
$\approx 10\%$ on the pulse amplitude, and therefore the
spin period of the NS in {\exo} remains undetermined.

\subsection{Type-I X-ray bursts}

Seven type-I X-ray bursts were observed during the 80~ks {\xmm}
observation presented here, with a recurrence time varying from 2.5 to
4 hours. None of the bursts showed photospheric radius expansion, and
all the bursts observed had a relatively long rise time ($\sim2$--5~s)
and decay timescale ($\tau=15$--$23$~s, except the second, brightest
burst which had $\tau\simeq10$~s). Bursts of pure helium are
characterized by shorter timescales ($\tau<10$~s) and we deduce that a
fraction of hydrogen was probably present in the fuel of the bursts we
observed.  More information on the fuel composition can be drawn from
the ratio between the integrated persistent flux and the burst
fluence, $\alpha$. This parameter is related to the ratio between the
efficiency of energy conversion through accretion onto a compact
object ($GM_*/R_*$) and thermonuclear burning
($Q_{\rm}=1.6+4$<X>~MeV~nucleon$^{-1}$, where <X> is the abundance of
hydrogen burnt in the burst), {$\alpha=44(Q_{\rm
    nuc}/4.4$~MeV~nucleon$^{-1})^{-1}$} for a 1.4~$M_{\odot}$ NS with
a radius of 10~km \citep[see Eq.~6 of][and references
  therein]{galloway2008}. The observed values of $\alpha$ range from
50 to 100, with an average of 82, indicating that hydrogen fraction in
the bursts was $<X>\approx0.2$. Mass accretion rate should have then
been high enough to allow stable hydrogen burning between bursts, but
part of the accreted hydrogen was left unburnt at the burst onset and
contributed to produce a longer event with respect to pure helium
bursts. Combined hydrogen-helium flashes are expected to occur for
mass accretion rates larger than $\simeq 0.1\,\dot{m}_{\rm Edd}$
\citep[for solar metallicity, lower values are expected for low
  metallicity,][]{2004ApJS..151...75W}, where $\dot{m}_{\rm Edd}$ is
the Eddington accretion rate per unit area on the NS surface
($8.8\times10^4$~g~cm$^{-2}$~s, or $1.3\times10^{-8}${\msunyear}
averaged over the surface of a NS with a radius of 10~km). The
persistent broadband X-ray luminosity of {\exo} during the
observations considered here indicates a mass accretion rate of
$8.5\times10^{-10}\,d_{5.5}^2\simeq0.05\,\dot{M}_{\rm Edd}${\msunyear}
for a 1.4~M$_{\odot}$ NS with a 10 km radius, lower than the above
threshold not to exhaust hydrogen before the burst onset. A low
metallicity could help decreasing the steady hydrogen burning rate and
leave a small fraction of hydrogen in the burst fuel.


The seven bursts observed during the {\xmm} observation analyzed here
share some of the properties of the 21 bursts observed by {\it RXTE}
during the 2000 outburst before the outburst peak, such as the decay
timescale, $\tau\approx25$~s, and the peak and persistent flux $F_{\rm
  peak}=(3$--$19)\times10^{-9}${\fluxcgs}, $F_{\rm
  pers}=(1$--$5)\times10^{-9}${\fluxcgs} and the absence of
photospheric radius expansion \citep[see Table 10 and appendix A31
  in][]{galloway2008}. However those bursts showed recurrence times
between 17 and 49 minutes, and correspondingly lower values of
$\alpha=20$--$46$ with respect to those observed here. The observation
of frequent, long bursts and infrequent, short bursts at similar X-ray
luminosity made \citet{galloway2008} classify {\exo} as an {\it
  anomalous} burster. The observations presented here confirm such
a puzzling behavior for {\exo}.
We note that 4 additional type-I bursts were detected by {\igr} during 
the monitoring observations of {\exo}.
As discussed in Sect.~2.2 we did not perform a spectroscopic analysis 
of these events due to the limited statistics of the two JEM-X units 
and the lack of any interesting detection in ISGRI which could have 
indicated the presence of a photospheric radius expansion phase.

\begin{acknowledgements}
Based on observations obtained with XMM-Newton, an ESA science mission with instruments and contributions directly funded by ESA Member States and NASA, and with INTEGRAL,  an ESA project with instruments and science data
centre funded by ESA member states, and Poland, and with the participation of Russia and the USA.
AP acknowledges  support via an EU Marie Sklodowska-Curie Individual Fellowship under contract No. 660657-TMSP-H2020-MSCA-IF-2014, as well as fruitful discussion with the international team on “The disk-magnetosphere interaction around transitional millisecond pulsars” at ISSI (International Space Science Institute), Bern. We also acknowledges financial support from INAF ASI contract I/037/12/0.

\end{acknowledgements}


\bibliographystyle{aa} 
\bibliography{exo1745}

\end{document}